\documentclass[preprint,authoryear,12pt]{elsarticle}

\usepackage{graphics}

\usepackage{float}
\usepackage{graphicx}
\usepackage{wrapfig}

\usepackage[draft]{todonotes}   

\usepackage[labelfont=bf]{caption}
\usepackage{caption}
\usepackage{color}
\usepackage{amsmath} 
\usepackage{mathtools} 
\usepackage{amsfonts} 
\usepackage{amssymb}
\usepackage{array,longtable}

\usepackage{booktabs}
\usepackage{rotating}
\usepackage{multirow}
\usepackage{multicol}
\usepackage{lscape}
\usepackage{subfig}
\usepackage{comment}
\usepackage{soul}

\usepackage{wrapfig}
\usepackage{rotating}
\usepackage{epstopdf}

\usepackage[utf8]{inputenc}
\usepackage[T1]{fontenc}
\usepackage{textcomp}
\usepackage{gensymb}

\usepackage[export]{adjustbox}

\usepackage{algorithm,algorithmicx,algpseudocode}

\usepackage{color}

\usepackage{comment}

\usepackage[colorlinks=true,linkcolor=blue,citecolor=blue]{hyperref}

\usepackage[draft]{todonotes}   

\usepackage{soul}

\definecolor{darkgreen}{rgb}{0.53, 0.66, 0.42}
\definecolor{azure}{rgb}{0.0, 0.5, 1.0}
\definecolor{carrotorange}{rgb}{0.93, 0.57, 0.13}
\definecolor{cornflowerblue}{rgb}{0.39, 0.58, 0.93}

\journal{Medical Image Analysis}

\begin{document}

\begin{frontmatter}



\title{Comparative Survey of Multigraph Integration Methods for Holistic Brain Connectivity Mapping}

\author[BASIRA,MANAGEMENT]{Nada Chaari}
\author[MANAGEMENT]{Hatice Camg\"{o}z Akda\u{g}}
\author[BASIRA,DUNDEE]{Islem Rekik\corref{cor}}

\cortext[cor]{Corresponding author: irekik@itu.edu.tr; \url{http://basira-lab.com/}.}

\address[BASIRA]{BASIRA lab, Faculty of Computer and Informatics, Istanbul Technical University, Istanbul, Turkey}
\address[MANAGEMENT]{Faculty of Management, Istanbul Technical University, Istanbul, Turkey}
\address[DUNDEE]{School of Science and Engineering, Computing, University of Dundee, UK \ }


\begin{abstract}

One of the greatest scientific challenges in network neuroscience is to create a representative map of a population of heterogeneous brain networks, which acts as a connectional fingerprint. The connectional brain template (CBT), also named network atlas, presents a powerful tool for capturing the most representative and discriminative traits of a given population while preserving its topological patterns. The idea of a CBT is to integrate a population of heterogeneous brain connectivity networks, derived from different neuroimaging modalities or brain views (e.g., structural and functional), into a unified holistic representation. Here we review current state-of-the-art methods designed to estimate well-centered and representative CBT for populations of single-view and multi-view brain networks. We start by reviewing each CBT learning method, then we introduce the evaluation measures to compare CBT representativeness of populations generated by single-view and multigraph integration methods, separately, based on the following criteria: centeredness, biomarker-reproducibility, node-level similarity, global-level similarity, and distance-based similarity. We demonstrate that the deep graph normalizer (DGN) method significantly outperforms other multi-graph and all single-view integration methods for estimating CBTs using a variety of healthy and disordered datasets in terms of centeredness, reproducibility (i.e., graph-derived biomarkers reproducibility that disentangle the typical from the atypical connectivity variability), and preserving the topological traits at both local and global graph-levels.

\end{abstract}

\begin{keyword}
Multiview brain connectivity \sep Multigraph integration \sep Connectional brain template \sep Graph fusion techniques

\end{keyword}

\end{frontmatter}


\section{Introduction}
 
The availability of large-scale neuroimaging datasets using non-invasive magnetic resonance imaging (MRI) has substantially increased our understanding of the extraordinarily complex, yet highly organized topology of the underlying human neural architecture; the so‐called connectome \citep{bullmore2009complex, fornito2015connectomics}. Using different sources of measurements, one can derive, for the same subject, multiple brain connectivity networks \citep{guan2020clinical}. Having such multimodal information, one can represent each subject by a multi-view graph where each view corresponds to an imaging modality quantifying a single type of brain connectivity network, each node of the graph denotes a brain region and the edge between two nodes represents the interaction between pairs of brain regions \citep{acosta2017extension, bunke2011recent}. A multigraph structure encapsulates the representation of multiple relations between two anatomical regions of interest (ROIs). For instance, connections in brain networks derived from resting-state functional magnetic resonance brain imaging (rsfMRI) encode correlations in functional activity among brain regions, whereas diffusion tensor imaging (DTI) networks provide information concerning structural connections (i.e., white matter fiber paths) between these nodes \citep{tyan2017gender, jiang2020gender,dadashkarimi2019mass}. Joining both networks results in two different views of brain connectivity, leading to more insights into the brain as a complex intersconnected system.

Understanding how the brain's structural, morphological, and functional levels interlink offers a more comprehensive picture of the brain facets construction \citep{bassett2017network}. However, analyzing such \emph{multi-modal} (also \emph{multi-view}) connectomic dataset together remains challenging due to the large inter-modality variations across different views of connectivity networks and the heterogeneity of brain networks across population samples \citep{van2016human, verma2019heterogeneous}. Nonetheless, mapping brain networks of a whole population into a single representation is substantial for capturing the most shared and representative brain connectivity patterns within a heterogeneous network population \citep{sporns2005human}. However, the brain connectivity map varies across individuals, which hampers the identification of neurological biomarkers in a specific disordered population. Mostly, such biomarkers are important in disentangling the typical from the atypical variations across the population individuals. For instance, numerous studies emphasized the importance of looking for commonalities and differences in neurobiological and psychiatric changes across brain disorders \citep{mahjoub2018brain, lisowska2019alzheimer, georges2020identifying}, which may improve our ability to understand the differences between comorbid disorders such as autism and dementia. Consequently, extracting an integral connectional fingerprint of heterogeneous brain networks of a given population while preserving their common and distinct patterns remains a critical pursuit towards the charting neurological disorder landscape \emph{at the population level} \citep{van2019cross}.

Human neuroscience studies have made significant progress on the path towards estimating brain network templates for a population of connectomes since the inception of the seed paper introducing the concept of `population-driven network atlas' in \citep{rekik2017estimation}. Several integration methods were proposed to form the integral representation for a population of unimodal (single-view) \citep{wang2014similarity, mhiri2020joint, mhiri2020supervised} as well as multi-modal (multi-view) \citep{dhifallah2019clustering,dhifallah2020estimation, chaari2020estimation, demir2020clustering, gurbuz2020deep} brain networks. Nevertheless, single-view integration methods were limited to fuse single-view networks, thereby, overlooking the complementary and richness of multi-view brain network populations. More broadly, multi-view integration methods generalized this concept to multi-view brain networks for more holistic and integral mapping of the brain connectivity population. Both categories of network fusion methods introduced the concept of connectional brain template (CBT) \citep{rekik2017estimation} as a normalized connectional representation of the population of single-view or multi-view brain connectomes while examining all population connectivities. Leveraging the brain template enables not only the integration of complementary information of a given connectomic population but also the generation of new connectomes for synthesizing brain graphs when minimal resources exist. A population-driven CBT can be used to guide brain graph classification as well as evolution prediction \citep{bessadok2021graph}. Furthermore, the estimation of a population CBT provides an excellent tool for extracting the integral connectional fingerprint of each population holding its most specific traits, which is an essential step for group comparison studies (e.g., investigating gender differences \citep{nebli2020gender}).

In this forward-looking review, we investigate and compare state-of-the-art single-view and multi-view integration methods, while focusing on how they produce for each population type (single-view-based and multiview-based networks, respectively) a unified normalized connectional representation (CBT). Most importantly, we conduct a comparative study between the unimodal fusion methods and the multimodal integration methods, separately, by evaluating the performance of their generated CBTs on a variety of healthy and disordered datasets in terms of (1) well-centeredness (2) discriminativeness, and (3) topological soundness to the population at different scales including node-wise similarity, distance-based similarity, and global-based similarity.  As these criteria are mandatory for a rigorous comparison between single-view or multigraph fusion methods, we used a combination of complementary evaluation methods and measures which investigate the soundness of the learned CBTs at different levels. Notably, we used four measures to assess the inter-method similarity at both graph-global and graph distance scales. For the graph-node scale, we extensively used 8 topological measures to assess the hubness behavior of the generated CBTs and we used 2 other topological metrics to evaluate their segregation and the integration behaviors.  Our extensive comparative study included 3-way evaluation strategies (\textbf{Fig.}~\ref{fig:AnalysisDiagram}): method-to-method analysis, ROI-based analysis, and method-based analysis. Combined together, these strategies enable the investigation of the centrality, the discriminability, and the topological strength preserved by the generated graphs (CBTs) when representing their populations.  We further discuss the results of the CBT evaluation measures and the strength of the best method. We highlight the limitations of the integration methods in estimating representative reference connectional templates derived from complex connectomes. We conclude with an outlook into the future of multigraph fusion methods and population-driven CBT learning techniques, identify unsolved challenges and discuss new avenues of improvement within this field. 

\begin{figure}[ht!]
\centering
\includegraphics[width=\linewidth]{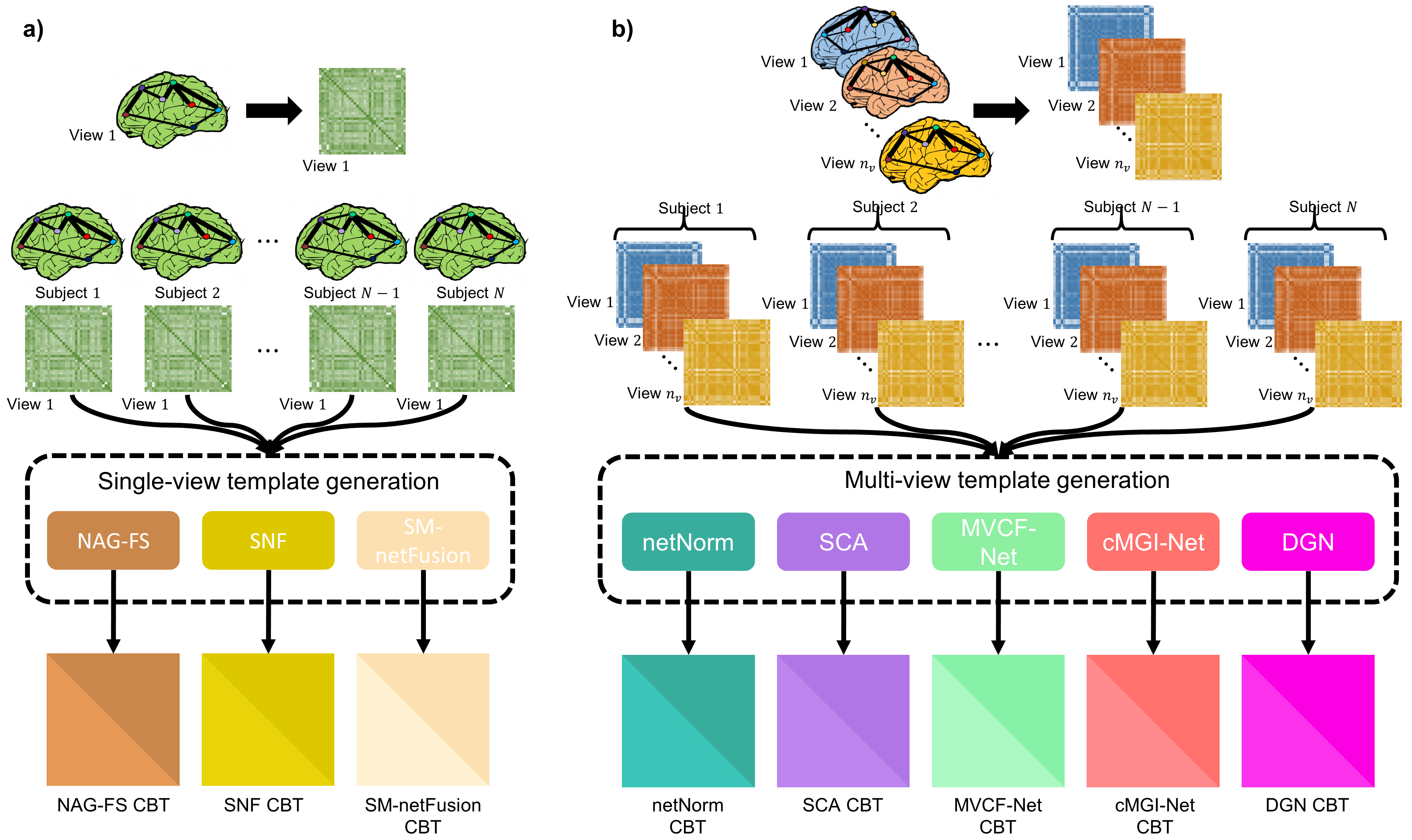}
\caption{The schema illustrates connectional brain templates (CBTs) estimated by \textbf{a)} single-graph fusion methods for a given population of single-view brain networks: network atlas-guided feature selection (NAG-FS) \citep{mhiri2020joint}, similarity network fusion (SNF) \citep{wang2014similarity} and supervised multi-topology network cross-diffusion (SM-netFusion) \citep{mhiri2020supervised}; and \textbf{b)} multi-graph integration methods for a population of multi-view brain connectivity dataset: multi-view networks normalizer (netNorm) \citep{dhifallah2020estimation}, cluster-based network fusion (SCA) \citep{dhifallah2019clustering}, multi-view clustering and fusion (MVCF-Net) \citep{chaari2020estimation} and cluster-based multi-graph integrator networks (cMGI-Net) \citep{gurbuz2020deep}.}
\label{fig:ConceptFigure}
\end{figure}

\section{Comparative Brain Multigraph Integration and Mapping Methods}

We present the first review paper which provides an insightful survey of the existing integration models promoted with a comparative study to evaluate their performance across extensive experiments in terms of producing the most centered templates, recapitulating unique traits of populations, and preserving the complex topology of biological networks. In our 2014-2020 search for articles that introduce graph integration methods, we emphasize two categories that estimate a unified connectional template of a population of networks. The first category corresponds to the single-view fusion methods where they take populations of single-view networks and output a single graph (or network). For this category, we identify three single-view graph fusion methods: SNF \citep{wang2014similarity}, NAG-FS \citep{mhiri2020joint}, and SM-netFusion \citep{mhiri2020supervised}. These network fusion methods are based on different machine learning (ML) architectures.

The second group represents multi-view graphs integration methods that fuse populations of multi-view networks into a single connectional template. For this category, we review five multigraph fusion methods: netNorm \citep{dhifallah2020estimation}, SCA \citep{dhifallah2019clustering}, MVCF-Net \citep{chaari2020estimation}, cMGI-Net \citep{demir2020clustering}, and DGN \citep{gurbuz2020deep}. Multigraph fusion methods can be sub-categorized into two big classes: \emph{machine learning-based} and \emph{deep learning-based} models. The source articles are published between 2018 and 2020, except the one proposed by \citep{wang2014similarity} in 2014. In the following section, we detail the architecture of each graph population integration method in each category. We refer the reader to our GitHub link where all papers cited in our work are available\footnote{\url{https://github.com/basiralab/survey-multigraph-integration-methods}}.

The aforementioned integration frameworks are graph-based models which are designed to learn from a brain graph where nodes represent anatomical brain regions, edges denote brain connectivities and edge weights represent connectivity weights. Integration models aim to generate a representative template graph which encodes the shared traits within a population of brain multigraphs (multi-view networks). Note that single-view integration is a special instance of the multi-view graph integration task where the number of views is 1. This problem can be defined as follows. Let sample $s$ (brain connectome) in a population be represented by a set of $n_v$ weighted undirected graphs, each comprising $n_r$ nodes. We model this sample as a single tensor $ \mathcal{T}_s \in \mathbb{R}^{n_r \times n_r \times n_v} $ that is composed of stacked $n_v$ adjacency matrices $ \{ \textbf{X}^{v}_{s} \}^{n_v}_{v=1}$ of $ \mathbb{R}^{n_r \times n_r} $capturing the pairwise relationships between $n$ ROIs. The objective of these fusion frameworks is to integrate a set of multi-view graphs $\emph{T} = \{ \mathcal{T}_1, \mathcal{T}_2, ..., \mathcal{T}_N \}$ in order to obtain a population-representative connectional template $ \mathcal{T} \in \mathbb{R}^{n_r \times n_r} $ that is inherently and by design: 

\begin{enumerate}
	\item well-centered (i.e., satistfies the minimal distance to all individuals in the populaiton),
	\item discriminative (i.e., capture distinctive population connectivities), and 
	\item topologically sound (i.e., the topological properties of the population individuals are preserved when transforming a population multi-view networks into a representative connectional template).
\end{enumerate}

\begin{figure}[ht!]
\centering
\includegraphics[width= 0.9\linewidth]{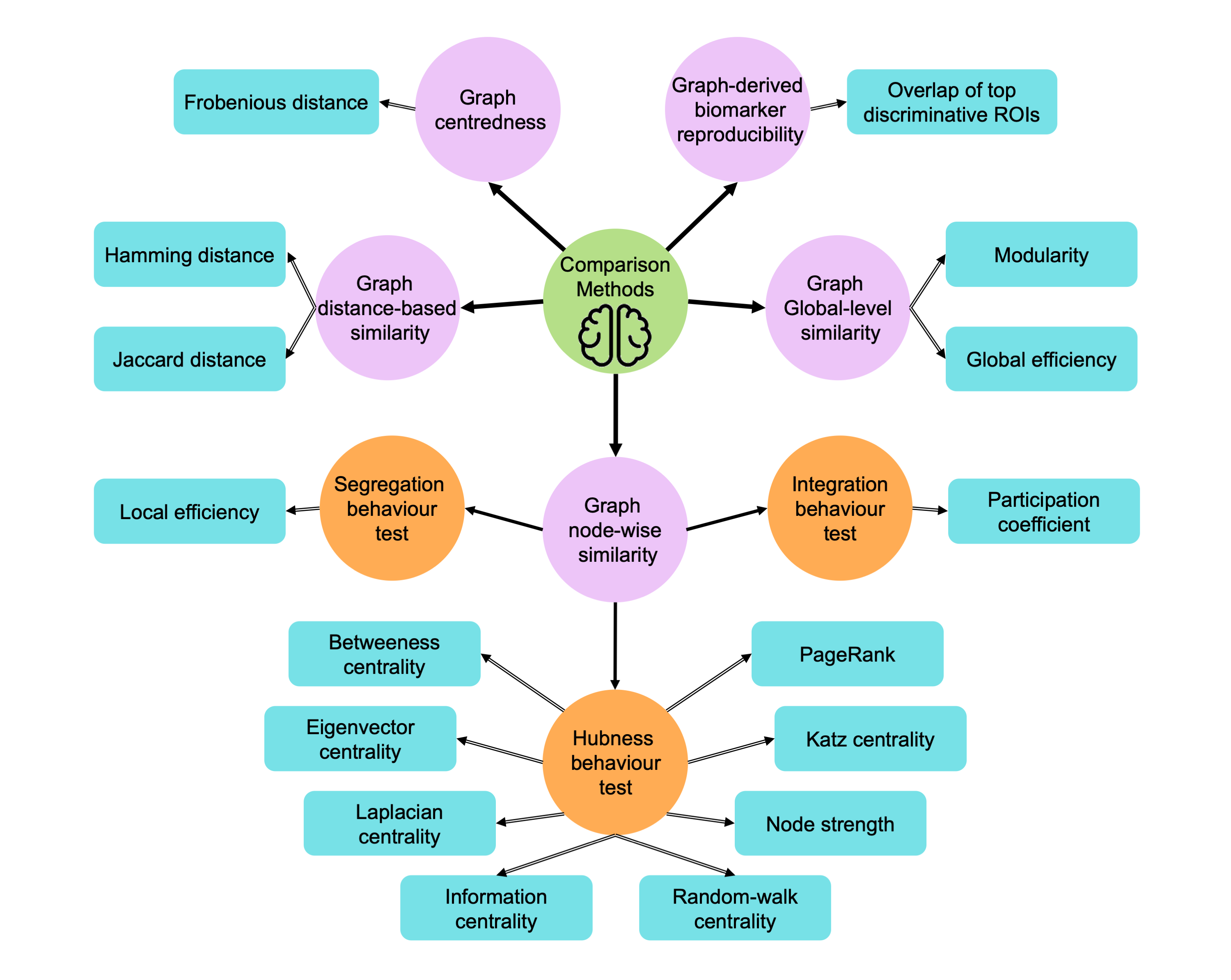}
\caption{Diagram illustrating the criteria used to evaluate the performance of the connectional brain templates generated for single-view graph integration methods and multi-view graphs fusion methods. This includes first graph centeredness using Frobenius distance, second, the graph-derived biomarker reproducibility where we identify the overlap of top $k$ discriminative nodes (ROIs) computed by the graph fusion methods and by an independent biomarker selection method, third the global-level similarity using both modularity and global efficiency measures, fourth the graph distance-based similarity where we compute Hamming and Jaccard distances. Finally, we include the graph node-wise similarity where we evaluate three behavior for the graph: integration behavior using participation coefficient, segregation behavior using local efficiency, and hubness behavior using the following centrality metrics: PageRank \citep{xing2004weighted}, Katz centrality \citep{katz1953new}, node strength \citep{barrat2004architecture}, random-walk centrality \citep{newman2005measure}, information centrality \citep{stephenson1989rethinking}, Laplacian centrality \citep{qi2012laplacian}, eigenvector centrality \citep{newman2008mathematics}, and betweeness centrality \citep{brandes2001faster}.}
\label{fig:ComparisonMethodsDiagram}
\end{figure}

\subsection{Single-graph fusion methods}

In this section, we cover in detail the three single-view integration methods that take as input a population of single graphs and output a unified CBT.

\textbf{\emph{SNF}}. Similarity network fusion (SNF), proposed by \citep{wang2014similarity}, is a generic unsupervised technique for non-linear network integration, which is based on message passing theory \citep{pearl2014probabilistic}. SNF aims to estimate a status matrix for each network that carries the whole network information and a sparse local matrix that only takes up to top-$k$ neighbors into consideration. Next, an iterative integration step is conducted to update each status network by diffusing the mean global structure of the remaining networks along with the sparse local network. This operation is iterated over each individual graph in the population. The network diffusion stops at an optimal convergence threshold where all diffused networks are linearly fused into a single network (i.e., connectional template).

\textbf{\emph{NAG-FS}}. The network atlas-guided feature selection (NAG-FS) method, proposed by \citep{mhiri2020joint}, is a feature selection-based method to produce a unified normalized connectional representation of a population of brain networks. First, NAG-FS clusters similar brain networks into non-overlapping subspaces using multiple kernels. Next, NAG-FS leverages the network diffusion and fusion techniques introduced in \citep{wang2014similarity} to nonlinearly fuse the networks lying in the same subspace, hence creating a cluster-specific network atlas. (i.e., a local population center). Last, the population connectional network atlas is obtained by non-linearly diffusing and fusing the cluster-specific network atlases. NAG-FS handles potential heterogeneity in the network population distribution at different bandwidths. 

\textbf{\emph{SM-netFusion}}. Supervised multi-topology network cross-diffusion (SM-netFusion), proposed recently by \citep{mhiri2020supervised}, is a supervised fusion method for CBT estimation from single-view networks of a population supervised by the data heterogeneity. SM-netFusion uses a weighted mixture of multi-topological measures to enhance the non-linear fusion for the proposed supervised graph integration. To the best of our knowledge, this work is a primer on \emph{supervised} graph population integration into a single graph. First, SM-netFusion learns a weighted combination of the topological diffusion kernels derived from the degree, closeness, and eigenvector centrality measures in a supervised manner. Next, SM-netFusion non-linearly cross-diffuses the normalized brain networks so that all diffused networks lie close to each other for the final fusion step to generate the target CBT of the input connectomic population. This normalization well captures the shared networks between individuals at different topological scales, improving the representativeness and centeredness of the learned multi-topology CBT.

\subsection{Multigraph fusion methods }

In this section, we detail the five state-of-the-art multi-view integration methods that are fed with a population of brain multigraphs, each capturing multiple types of interactions between brain regions, and overview how these techniques generate the holistic CBT.

\textbf{\emph{SCA}}. Starting with the cluster-based network fusion (SCA) introduced by \citep{dhifallah2019clustering}, the proposed method is a multi-view brain connectivity fusion framework for estimating a brain network atlas from a multi-view brain network population. Specifically, SCA non-linearly fuses multi-view networks into a single network for each subject. In this step, all individuals are first mapped from the original space into the mapped space where their brain views are unified individually by leveraging the generic SNF \citep{wang2014similarity}. Next, SCA clusters the fused networks in the mapped space to identify individuals sharing similar connectional traits in an unsupervised way, which are next averaged within each cluster to generate a representative network atlas for each cluster. After obtaining the cluster-based brain templates, SCA constructs the final multi-view network atlas by linearly averaging the obtained templates of all clusters into a single template denoting the holistic network atlas.

\textbf{\emph{netNorm}}. As a more advanced fusion technique, netNorm was introduced by \citep{dhifallah2020estimation}, building a representative template based on a novel graph feature selection technique prior to a non-linear fusion for the target multi-view network integration. First, netNorm defines a cross-view feature vector between each pair of ROIs for each individual in the population. In order to investigate the inter-relationship between different subjects in a population at a local scale, this framework constructs a high-order graph for each pairwise connection by measuring the Euclidean distance between the cross-view feature vectors across all subjects. Next, netNorm selects the most centered cross-view connectional features (i.e., connectivity weights) across the population individuals, which pins down the most representative connectivities for each network view. Finally, the network views are integrated into a single network using non-linear fusion technique to output the final brain connectional template.

\textbf{\emph{MVCF-Net}}. More recently, \citep{chaari2020estimation} proposed multi-view clustering and fusion (MVCF-Net), a graph-based clustering method, to fuse a population of multi-view networks. This method is rooted in the identification of consistent and differential clusters across brain views to generate a connectional brain template for a given population. To this aim, first, MVCF-Net leverages a multi-view network clustering method based on manifold learning \citep{yu2019simultaneous}, which groups similar subjects in the same cluster and separates dissimilar subjects in different clusters while preserving their alignment across data views. Thus, similar connectional traits and distinct connectional traits of graphs within and between clusters across various data views can be identified in a fully unsupervised way \citep{yu2019simultaneous}. Next, for each view, MVCF-Net linearly fuses the networks within each cluster to generate a local CBT and non-linearly integrates the resulting local CBTs across views into a cluster-specific CBT. Next, by linearly fusing the cluster-specific centers, the final CBT is estimated to represent a given population of multi-graph networks. MVCF-Net jointly captures simultaneously similar and distinct connectional traits of samples. 

\textbf{\emph{cMGI-Net}}. More recently, \citep{demir2020clustering} proposed a clustering-based multi-graph integrator network (cMGI-Net) for CBT estimation of multigraph population. Based on geometric deep learning, cMGI-Net non-linearly maps a population of brain multigraphs to a target CBT in an end-to-end manner using a single objective loss function to optimize. First, cMGI-Net clusters similar samples together using multi-kernel manifold learning (MKML) introduced in \citep{wang2018simlr} to disentangle the heterogeneous distribution modes of the given graph population, thereby facilitating the following integration task. Next, for each cluster, cMGI-Net integrates the multigraph network of each subject into a single subject-specific graph to identify useful edge types between connected nodes. This step results in generating metapaths across-views for each subject which can encapsulate representative connectivities within a multigraph network. Next, cMGI-Net fuses the generated subject-specific graphs into a cluster-specific CBT while learning their weights under the constraint of minimizing the distance between the resulting template and all multigraph networks in the population. The final CBT is estimated by simply averaging the cluster-specific CBTs.

\textbf{\emph{DGN}}. Another very recent approach, deep graph normalizer (DGN) introduced by \citep{gurbuz2020deep}, is a graph neural network (GNN) architecture that learns how to normalize and integrate a population of multigraph brain networks into a single CBT in \emph{end-to-end} manner. First, each training sample passes through a sequence of graph convolutional neural network layers which are separated by non-linear activation of the previous layer. Precisely, each GNN layer learns deeper embeddings for each node by locally integrating connectivities offered by different heterogeneous views and blending the previous layer's embeddings using integrated connectivities. Next, DGN computes the pairwise absolute difference of each pair of the final layer's node embeddings to derive connectivity weights of the generated CBT. To evaluate the representativeness of the estimated subject-biased CBT, DGN is based on a randomized subject normalization loss(SNL) which updates the model weights for each subject in a way that the learned subject-specific CBT is representative of a random subset of the training subjects. Specifically, the trained model is fed with an arbitrary subject of the training population and learns how to achieve subject-to-population mapping thanks to SNL optimization. The final population CBT is generated by selecting the element-wise median of all training subject-specific CBTs, thereby retaining the most centered connections.

\section{Comparative Experiments and Evaluation Measures}

In this section, we  structure our comparative experimental design to evaluate single-view and multigraph population integration methods for brain connectional template estimation into 3-way evaluation strategies (\textbf{Fig.}~\ref{fig:AnalysisDiagram}). \emph{First,} we conduct ROIs-based analysis where we extensively assess the CBTs generated by the integration methods against the ``ground-truth'' population at the node-graph scale using 8 topological measures, local efficiency, and participation coefficient metrics. These metrics allow an extensive investigation about how much the generated CBTs are similar to the population to represent at the node level. \emph{Second,} we conduct method-based analysis to evaluate the estimated CBTs at the global-graph level using the average of the pre-mentioned 8 topological measures over the ROIs, Frobenius distance, Kullback-Leibler divergence, graph modularity, local efficiency, and global efficiency metrics. \emph{Third,} we conduct method-to-method analysis to investigate the similarity in performance between pairs of graph integration methods using KL-divergence, Hamming distance, and Jaccard distance measures. Together, these strategies provide a rigorous comparison between integration methods, and thus a reliable evaluation of the learned CBTs in terms of centeredness, discriminativeness, and preserving the topological soundness at the node-level, global-level, and distance-based level.

Ideally, a reliable learned CBT should preserve the topological patterns and properties of individuals in the input population during the integration process \citep{bullmore2009complex}. Specifically, a CBT should satisfy the following criteria: 

\begin{enumerate}
	\item Centeredness as it occupies the ‘center' of a population by achieving the minimum distance to all population samples. 
	\item Graph-derived biomarker reproducibility as it allows to identify connectional biomarkers that disentangle the differences in brain connectivity between populations with different brain states (i.e., healthy and disordered or genders).
	\item Graph global-level similarity as it tests whether the generated CBT preserves the global structure of the original graphs networks. 
	\item Graph node-wise similarity as it tests whether the local structure of the original data which includes the relationship (connectivity) between the nodes is preserved by the CBT.
	\item Graph distance-based similarity as it quantifies the distance between two networks based on predefined similarity scores.
\end{enumerate}

To evaluate the centeredness of the CBTs, we measure the Frobenius distance from the estimated template to each brain tensor view of each subject in the population. To rigorously compare different graph or multigraph fusion methods, it is necessary to quantify the similarity between them \citep{huang2018graph, yu2018application} using graph theoretical measures. However, relying on one topological measure might be reductionist given the convoluted connectivity of the brain as a network. To overcome such flaw in our experimental design, we propose a combination of complementary evaluation methods and metrics that examine the learned CBT at different levels: (i) the nodal level (brain regions), (ii) the local level, and (iii) the global level of a brain network \citep{christmas1995structural, luo2001structural, mheich2017siminet, shimada2016graph, wilson2008study}. In the rest of this section, we detail the five criteria for comparing the reviewed CBT learning frameworks, their evaluation methods, and measures.

\subsection{Cross-validation and CBT centeredness test}

We evaluate the centeredness and representativeness of the estimated CBT by measuring the mean Frobenius distance from the estimated template to each tensor view of each subject for a given population. Frobenius distance between two matrices $\mathbf{A}$ and $\mathbf{B}$ is a scalar value and is calculated as: $d_F (\textbf A,\textbf B) = \sqrt{\sum_{i}\sum_{j} \rvert{\textbf A_{ij} - \textbf B_{ij}}\rvert^2 }$. For reproducibility and generalizability, we split the datasets into training and testing sets using 5-fold cross-validation. We use the training set to generate a connectional template and then calculate its mean Frobenius distance to all views of each subject in the left-out testing fold. Hence, for each input dataset, we generate 5 CBTs, with an additional one using the whole data samples. To assess the statistical significance of each single-view fusion method and multigraph fusion method, we validate the comparative study of the CBT centeredness using a two-tailed paired t-test across all data folds in addition to the whole data between the comparative methods.

\subsection{CBT discriminativeness reproducibility test} 

Our second criterion is that the generated population templates are discriminative which implies that CBTs encapsulate the most distinctive traits of a population of graph networks. To test such hypothesis, we first spot the most $k$ discriminative brain ROIs where a population $p^A$ CBT largely differing from a population $p^B$ CBT (e.g., (1) Alzheimers' disease vs late mild cognitive impairment patients and (2) male vs female). To do so, we compute the absolute difference between both estimated templates $p^A$ and $p^B$, respectively. To assess the reproducibility of the CBT produced by each graph fusion method, we use randomized $k$-fold partition to divide each population into $k$ folds. $\textbf A_i$ and $ \textbf B_j$ denote the estimated CBTs for the $i^{th}$ fold for population $p^A$ and the $j^{th}$ fold for population $p^B$, respectively, where $1 \leq i,j\leq k$. We compute the mean absolute difference between the estimated templates across folds using simple element-wise inter-template subtraction as follows: $\textbf D = \sum_{i,j}\rvert{\textbf A_{ij} - \textbf B_{ij}}\rvert , 1 \leq i,j\leq k$, where $\textbf D$ is an $n_r \times n_r$ matrix containing the absolute features' differences between all pairs of fold $i$ and fold $j$ in connectivity weights. Next, we sum the columns of the resulting difference matrix to obtain a discriminability score vector $\alpha$ where the $i^{th}$ coefficient denotes the score $\alpha_i$ assigned to the $i^{th}$ ROI representing the cumulative distance from ROI $i$ to all other ROIs $k \neq i$. $\alpha_i$ is calculated as follows: $\alpha_i = \sum_{k} \textbf D(i,k), 1 \leq k\leq n_r, k \neq i$. We then pick the top $k$ discriminative ROIs with the highest scores. \\

To evaluate the reproducibility of CBT-driven discriminative ROIs, we propose to use an independent learner, namely multiple kernel learning (MKL) \citep{varma2009more}, which aims to identify the most discriminative features for a target classification task disentangling both $p^A$ and $p^B$ groups. Next, we compute the overlap (in \%) between the top discriminative ROIs found by (i) the mean absolute difference between the estimated CBTs and (ii) a supervised machine learning method based on MKL, respectively. To do so, we independently train a support vector machine (SVM) based on a supervised feature selection method. For each network view $v$, we first extract connectivity weights from each brain network view belonging to the given population by vectorizing the upper triangle of its connectivity matrix. Next, we use a $k$-fold randomized partition to divide each population $p^A$ and $p^B$ into $k$ sub-populations. Given the $v^{th}$ brain view, for each combination of $p^A_i$ and $p^B_i$ sub-populations, where $1 \leq i,j \leq k$ , we train an SVM classifier using the wrapper MKL method to learn a weight score vector quantifying the importance of each brain connectivity according to their distinctiveness in distinguishing between two subpopulations with different brain states (e.g., healthy vs disordered ). Next, we compute the total feature weight vector by summing up the weight vectors across all views and all possible $(\textbf A, \textbf B)$ combinations of $k$ sub-populations as follows: $ \textbf w = \sum_{v=1}^{n_v}\sum_{i,j=1}^k w_{i,j}^v$. We linearly anti-vectorize the resulted feature weight vector $w$ to obtain matrix $\textbf M \in \mathbb{R}^{n_r \times n_r}$, where each element $\textbf M(i, j)$ represents the learned weight assigned to a brain connection between ROIs $R_i$ and $R_j$. Next, we sum up the columns of the resulting matrix $M$ to obtain ROI-based discriminability scores where each weighted score $\alpha_i$ quantifies the discriminative power of an ROI $R_i$. Finally, we pick the top ROIs with the $k$ highest score. Following the identification of the top $k$ discriminative ROIs by each CBT-based graph fusion method and the MKL-based SVM method, respectively, we report their overlap. The CBT learning method achieving the highest overlap is best in \emph{reproducing} the most discriminative ROIs.
\subsection{CBT node-wise similarity comparison}

Many studies, which investigate the topological features of complex networks \citep{bullmore2009complex, guimera2005worldwide, watts1998collective}, consider that graph theoretical measures are sufficient to preserve the population topology. Among them, we specify graph node-wise measures which can be estimated at the node level of the compared networks. Graph node-wise measures are calculated for each node, and then the node's measure values are compared across brain CBTs. Such comparison measures allow not only to explore more features in the graph, but also indicate where the difference is located between the CBTs (e.g., which brain regions differ). Complementary topological measures need to investigate three main behaviors in a given brain network: hubness, segregation, and integration \citep{sporns2013network,cohen2016segregation}.

\subsubsection{CBT hubness behaviour test}

\begin{figure}[ht!]
\centering
\includegraphics[width=1\linewidth]{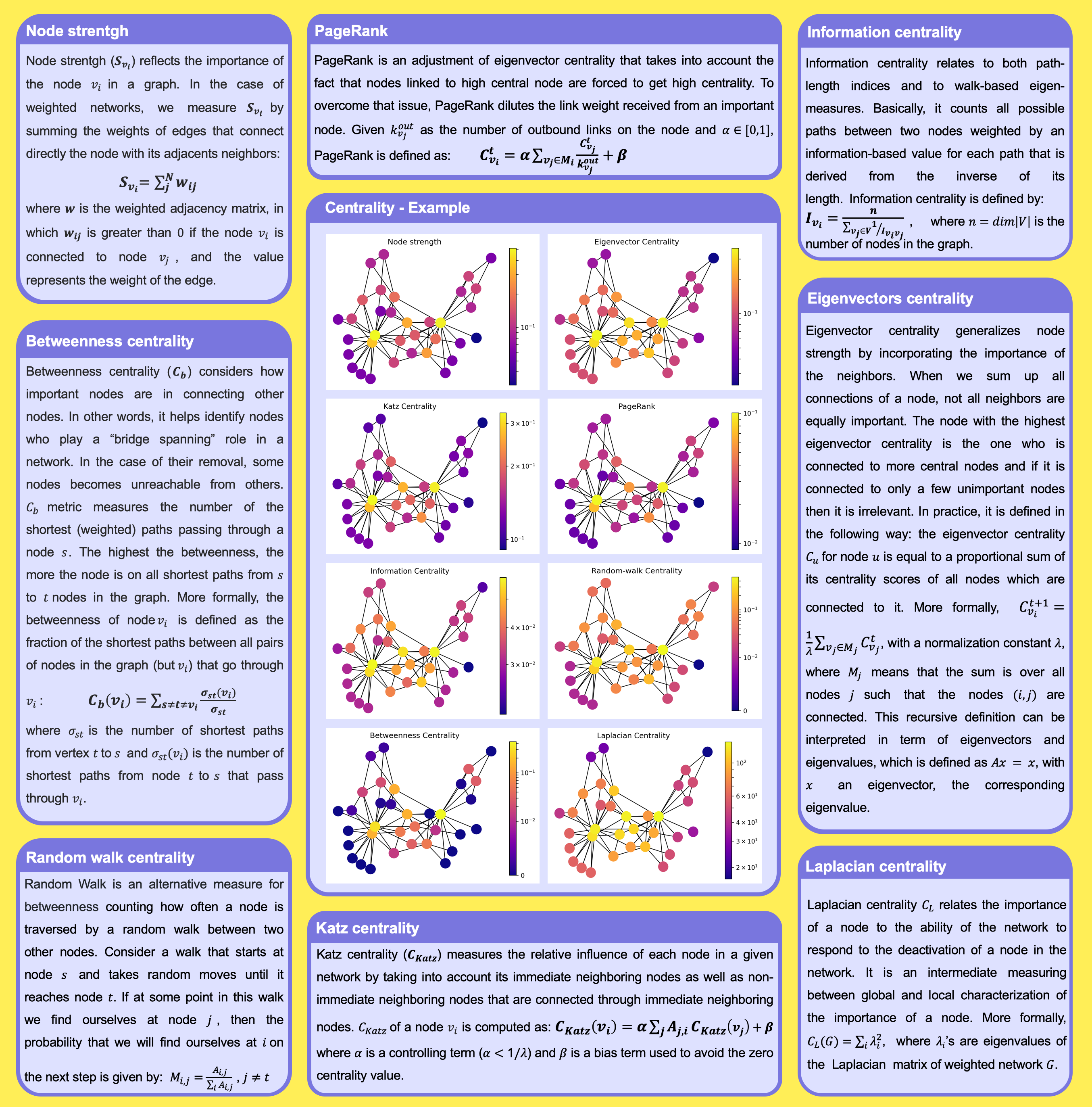}
\caption{\emph{Hubness centrality measures used for comparing CBTs generated by the reviewed methods}.}
\label{fig:HubnessMeasures}
\end{figure}

To evaluate the hubness behavior of population-driven CBT (graph), we first include the following measures as they capture different graph topological properties (\textbf{Fig.}~\ref{fig:HubnessMeasures}): (i) node strength \citep{barrat2004architecture} sums the connectivity weights of a particular node to all other nodes, (ii) betweenness centrality \citep{brandes2001faster} is defined as the fraction of all shortest paths in the network that pass through a given node, (iii) random-walk betweeness centrality \citep{newman2005measure} counts how often a node is traversed by a random walker between two other node, (iv) eigenvector centrality \citep{newman2008mathematics} measures a node's importance while giving consideration to the importance of its neighbors, (v) weighted PageRank \citep{xing2004weighted} can be seen as a variant of eigenvector as it investigates the in-degree of nodes and their neighbours by assigning a score to each node based on the number and the weight of edges connected to each node, (vi) Katz centrality \citep{katz1953new} can be seen as the generalization of the eigenvector centrality where it computes the relative influence of a node within a network by measuring the number of immediate neighbors (first degree nodes) and also all other nodes in a network that connect to a node under consideration through these immediate neighbors, (vii) information centrality \citep{stephenson1989rethinking} is defined as a variant of closeness centrality based on effective resistance between nodes in a network --quantifying how easy a node is reached by paths from other nodes--, and (viii) Laplacian centrality \citep{qi2012laplacian} quantifies each node using node strength to asses the impact of their removal from a graph.

Next, we conduct an ROI-based comparison using the distribution over the ROIs of the aforementioned centrality measures (CMs) for the learned connectional templates (CBTs-based CMs) generated by the reviewed single-view and multi-graph integration methods, separately. Specifically, each CM distribution is a discrete distribution that is composed of topological measures calculated for each node. For a fair comparison between the CBT-based CMs, we include CM distribution of the ground truth template (GT-based CM), as a reference for the estimated templates. For each centrality measure, we acquire the ground truth distribution by averaging the distribution of topological measures (i.e., PageRank) of each network view of each testing subject.

Additionally, we compute method-based comparison using the Kullback-Liebler divergence measure (KL-divergence). For each centrality measure and for each graph fusion method, we compute the KL-divergence between the CBT-based CM distribution and the GT-based CM distribution over brain ROIs. Note that we normalize each distribution using the total sum of measures across all nodes before computing KL-divergence to get a valid discrete probability CM distribution. 

For the reproducibility of hubness results, we apply 5-fold cross-validation to split the data into a training set used for CBT generation and a testing set. We first compute the topological distributions of the training connectional templates generated by single-view and multigraph fusion methods and the centrality measure distribution of the ground-truth test population, and then to compute the KL-divergence between the GT-based CM distribution and the CBTs-based CM distributions. We report the average centrality measure distributions across folds for the learned CBT as well as the average of the ground truth distributions. We further compute the average Kullback-Liebler divergence over folds between the normalized CBT-based CM distribution and the normalized GT-based CM distribution. More formally, KL-divergence measures the difference between two probability distributions $p(x)$ and $q(x)$ of a discrete variable $x$:

\begin{equation}
  D_{KL}(p(x) \parallel q(x)) = \sum_{x=1}^{n_{v}} p(x) \ln \frac {p(x)} {q(x)},  
\end{equation}

where $p$ denotes the normalized CBT-based CM distribution, $q$ is the normalized GT-based CM distribution and $x$ represents a particular node (ROI) in the brain graph. Finally, we conduct a method-to-method comparison between all possible pairs of CBT estimation single-view and multigraph methods. Specifically, we average the centrality measure distributions across folds for the CBT learned by each fusion method. Next, we compute the KL-divergence between all pair methods using their normalized CBT-based CM distribution, while considering all possible pair combinations.

\begin{figure}[ht!]
\centering
\includegraphics[width=0.5\linewidth]{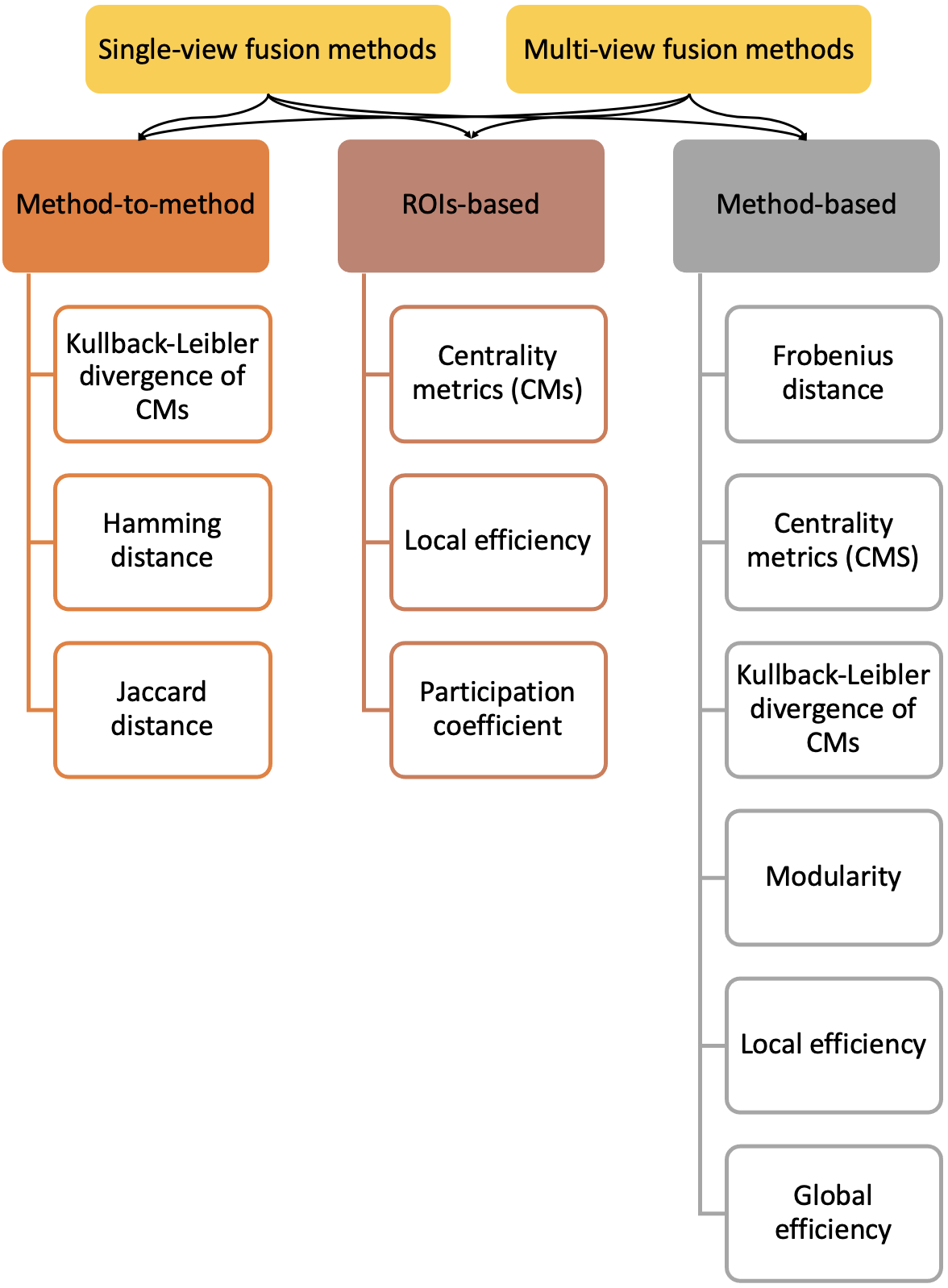}
\caption{Analysis diagram illustrating the types of comparative study to evaluate single-view and multigraph population integration methods for brain connectional template (CBT) estimation. We conduct method-based analysis where we evaluate the CBTs generated by the fusion methods against the ground truth using Frobenius distance, centrality metrics, Kullback-Leibler divergence, modularity, local efficiency, and global efficiency measures. We also conduct ROIs-based analysis to evaluate the CBTs at the node-scale, where we recompute both centrality metrics and local efficiency, and we add the participation coefficient. Lastly, we include method-to-method analysis using a pairwise comparison between graph fusion methods, where we compute three measures: KL-divergence, Hamming distance, and Jaccard distance.}
\label{fig:AnalysisDiagram}
\end{figure}

\subsubsection{CBT segregation behaviour test}

By comparing CBTs generated by the different methods at the node level, we evaluate the segregation behavior of the learned population template, quantifying the relative difference in the strength of within-network versus between-network brain connections. Among the measures belonging to this class, we include the local efficiency \citep{latora2001efficient}, which measures the efficiency of information transfer limited to neighboring nodes. It is calculated as the average nodal efficiency among the neighboring nodes of node $i$, excluding node $i$ itself (\textbf{Fig.}~\ref{fig:HubnessMeasures}). More formally, the local efficiency is defined as:

\begin{equation}
  E_{loc}(i)= \frac{1}{N_{G_i}(N_{G_i}-1)} \sum_{j \neq h \in G_i}{\textbf A_{j,h}}
\end{equation}

where $\textbf A_{j,h}$ is the connectivity weight of the adjacency matrix $\textbf A$ of graph $G$ linking node $j$ with node $h$, and $N_{G_i}$ is the number of neighbour of node $i$.


To assess the segregation behavior of each population-driven CBT, we compute the local efficiency distribution across the brain ROIs so that we investigate how much information each ROI (node) transmits to its neighbors. For a fair comparison between the integration methods and to evaluate the performance of the generated CBTs based on this measure, we further compute the local efficiency distribution of the ground truth which represents the test population by averaging the local efficiency calculated for each view  of each subject in the population, independently. The CBT showing striking similarity distribution with the ground truth (local efficiency measure), achieves the highest performance in terms of preserving the segregation topological behavior. For easy visualization of the evaluation, we display the average local efficiency distribution across the ROIs of each for the learned CBTs and for the ground truth brain networks (\textbf{Fig.}~\ref{fig:HubnessMeasures}).

\subsubsection{CBT integration behaviour test}

Another way to measure the similarity between graphs is to evaluate their integration behavior, which reflects the ability of the network to combine information from distant nodes. In this context, we include participation coefficient \citep{guimera2005worldwide}, which quantifies the balance between the intra-module versus inter-module connectivity for a given node. In other words, the participation coefficient measures the distribution of a node's edges among the communities of a graph. This metric approaches 0 when a node's edges are restricted to its own community, and it takes a maximal value that approaches 1 when the node's edges are equally distributed among all communities (high correlations with multiple communities). More formally, the participation coefficient can be defined as:
\begin{equation}
    PC_i = 1- \sum_{s=1}^{N_c}\frac{k_{i,s}}{k_i}
\end{equation}
where $k_{i,s}$ is the node strength (sum of weights connections) of node $i$ to other nodes in its own community network $(s)$, and $k_i$ is the degree of node $i$ regardless of community membership. By subtracting that ratio from 1, the participation coefficient becomes a normalized measure of the connections that are not within a node's own community, or that are across communities.


To assess the integration behavior of each estimated population-driven CBT, we compute for each connectional brain template its participation coefficient to investigate the ability of CBT in combining information from its communities. We similarly quantify the participation coefficient of the ground truth by averaging the participation coefficient calculated for each view of each subject in the population, independently. The CBT achieving the highest global efficiency score shows the highest performance by preserving the data topology in terms of integration behavior.

\subsection{CBT graph-edit distance-based comparison}

The main idea of distance-based graph comparison methods consists of comparing two graphs by quantifying their similarity. This includes methods based on graph edit distances that focus on finding the common/uncommon nodes (brain regions) and edges (connections) between two brain networks. A special instance of the broader class of graph-edit distance is the Hamming distance. Introduced by \citep{gao2010survey}, the Hamming distance measures the amount of change between two graphs by counting the number of edge deletions and insertions necessary to fully transform one graph into another. More formally, the (normalized) Hamming distance is defined as the sum of the difference between the adjacency matrices of two graph networks $G$ and $\tilde{G}$ on $N$ nodes:

\begin{equation}
   d_H(G,\tilde{G})= \sum_{i,j=1}^N \frac{\left |\textbf A_{i,j} -\tilde{\textbf A}_{i,j} \right |}{N(N-1)}= \frac{1}{N(N-1)}\left \| \textbf A-\tilde{\textbf A} \right \|_{1,1}
\end{equation}

where $i$ and $j$ are two nodes, and $\mathbf{A}$ and $\tilde{\mathbf{A}}$ are the adjacency matrices of graphs $G$ and $\tilde{G}$ , respectively. The Hamming distance value (if normalized) is bounded between 0 (no similarity at all) and 1 (fully similar/same network) overall graphs of size $N$. However, the Hamming distance is sensitive to the density of the graphs. This yields a limited capacity to recognize a similar level of relative variability across graphs with varying sparsity. 

A potential solution to the aforementioned density-effect problem consists in using the Jaccard distance \citep{levandowsky1971distance}, which includes a normalization with respect to the volume of the union graph. This distance metric can be understood as the proportion of edges that have been removed or added with respect to the total number of edges appearing in either graph network. More formally, given two weighted graphs, $G$ and $\tilde{G}$, and two nodes $i$ and $j$, the Jaccard similarity is defined as the difference between the size of the intersection of a graph $G$ and graph $\tilde{G} $ (i.e. the number of common edges) and the size of the union of a graph $G$ and graph $\tilde{G} $ (i.e. the number of unique edges) over the size of the union of a graph $G$ and graph $\tilde{G} $: 
\begin{equation}
   d_{Jaccard}(G,\tilde{G})= \frac{\left | G \cap\tilde{G} \right |\left | G \cup \tilde{G} \right |}{\left | G \cup \tilde{G} \right |}
\end{equation}
In the case of weighted graphs, $G$ and $\tilde{G}$ can be represented by their corresponding adjacency matrices $\textbf A$ and $\tilde{\textbf A}$, where $ \textbf A_{i,j}$ and $\tilde{\textbf A}_{i,j}$ denote the edges weights of graphs $G$ and $\tilde{G}$, respectively, connecting node $i$ and node $j$. The Jaccard distance can be formalized as: 
\begin{equation}
   d_{Jaccard}(G,\tilde{G})=1-\frac{\sum_{i,j}min(\textbf A_{i,j},\tilde{\textbf A}_{i,j})}{\sum_{i,j}max(\textbf A_{i,j},\tilde{\textbf A}_{i,j})}
\end{equation}

A Jaccard distance close to 1 indicates an entire remodeling of the graph structure between graph $G$ and $\tilde{G}$.


To assess the graph-edit distance-based comparison between single-view or multigraph integration methods, we compute the proposed distance measures between all possible pairs of generated CBTs so that we evaluate their distance-based similarity in terms of sharing common brain regions and connections. The output scores indicate how much the CBT pairs are dissimilar.

\subsection{CBT global-level similarity comparison}

In this part, we aim to evaluate the global structure of the estimated CBTs. One way is to investigate the modular structure of the learned CBT by a particular method. Here, we adopt the modularity definition $Q$ introduced by \citep{newman2004finding}, which evaluates the goodness of partitioning of graph nodes into clusters. In other words, the modularity detects the communities (clusters) in a graph where a node belongs to a community if it has stronger connections with members of this community than with members of another community. Thus, high modularity indicates a good clustering where dense connections between nodes are within the same cluster and sparse connections are in different clusters, whereas low modularity entails poor clustering. More formally, given an adjacency matrix $\textbf A\in \mathbb{R}^{n_r\times n_r} $ which represents the estimated connectional template (CBT) in our case, the modularity $Q$ applies to a graph $G$ and a clustering $C$ can be written as:    
 
\begin{equation}
Q\left ( C \right )= \sum_{i,j\in V} \left (\textbf A_{i,j}-\frac{w_i w_j}{w} \right )\delta_C\left ( i,j \right ),
\end{equation}

\begin{equation}
    \begin{cases}
      w_i = \sum_{i\in V} \textbf{A}_{i,j} \\ 
      w= \sum_{i\in V} w_i= \sum_{i,j\in V} \textbf{A}_{i,j} \\
      \delta_C\left ( i,j \right ) =1, \text{if $(i,j)$ are in the same cluster under clustering $C$} \\ 
      \delta_C\left( i,j \right)=0,   \text{otherwise.}  
    \end{cases}       
\end{equation}

where $\textbf A_{i,j}$ denotes the connection weight that relates node $i$ with node $j$. A minimum value of $Q$ near to $0$ indicates that the considered network is close to a random one, whereas a maximum value of $Q$ near to $1$ indicates a strong community structure. The modularity $Q$ can be written in terms of probability distribution:

\begin{equation}
  Q\left ( C \right )= \sum_{i,j\in V} \left ( p\left ( i,j \right )-p\left ( i \right )p\left ( j \right )\right )\delta_C\left ( i,j \right ).
\end{equation}

Our objective is to cluster nodes while maximizing the modularity which means to decrease the second term of $Q$. However, this quantity is negligible for too small clusters. To go beyond the resolution limit, the multiplicative factor $\gamma$, called the resolution is introduced as follow:

\begin{equation}
    Q\left ( C \right )= \sum_{i,j\in V} \left ( p\left ( i,j \right )-\gamma  p\left ( i \right )p\left ( j \right )\right)\delta_C\left ( i,j \right ).
\end{equation}


To quantify the modularity of the learned CBT, we leverage the hierarchical clustering algorithm, named ``Pairwise Agglomerative using Resolution Incremental sliding'' (Paris) \citep{bonald2018hierarchical, newman2006modularity}. The main idea of this method is to split optimally the nodes of a network into $K$ non-overlapping communities using the first value of the resolution parameter $\gamma$, say $\gamma_1$, which can be written as:

\begin{equation}
    \gamma_1= max_{i,j\in V}\frac{p(i,j)}{p(i)p(j)}.
\end{equation} 

A second method to evaluate the entire structure of the estimated CBTs is to quantify the exchange of information across the whole graph network. A good evaluation measure for this property is the global efficiency \citep{latora2001efficient, achard2007efficiency}, which is defined as the inverse of the average distance (efficiencies) over all pairs of nodes $(i,j), i \neq j$ in the whole graph. More formally, the global efficiency is denoted:  

\begin{equation}
    E_{glob} = \frac{1}{n (n-1)}\sum_{i \neq j} \frac{1}{d(i,j)}
\end{equation}


To assess the similarity of the learned CBTs on the graph global-scale, we quantify for each connectional brain template its global efficiency as it is a more biologically relevant measure to capture information flow in brain networks \citep{latora2001efficient, bassett2006small}. The CBT showing the highest performance in terms of preserving the global topology of the data (population graph networks), achieves the highest global efficiency score.

\section{Results and Insights}

\subsection{Datasets}

We conduct our comparison study between different CBT estimation methods using the aforementioned evaluation measures on two multi-view connectomic datasets: the first dataset (M/F dataset) consists of 308 male subjects (M) and 391 female subjects (F) from the Brain Genomics Superstruct Project (GSP) dataset \citep{holmes2015brain}, aged between 21 and 23 years old; males ($n$ = 308; 21.6 ± 0.9 years, mean ± s.d.); females ($n = 390, 21.6 \pm 0.8$ years, mean $\pm$ s.d.). The second dataset (AD/LCMI dataset) is collected from the Alzheimer's Disease Neuroimaging Initiative (ADNI) database GO public dataset \citep{weiner2015impact} and includes 67 subjects (35 diagnosed with Alzheimer's diseases (AD) and 32 with Late Mild Cognitive Impairment (LMCI)).  For both datasets, each subject is represented by 4 cortical morphological brain networks \citep{nebli2020gender} derived from maximum principal curvature, mean cortical thickness, mean sulcal depth, and average curvature measurements. For each hemisphere, the cortical surface is reconstructed from T1-weighted MRI using FreeSurfer pipeline \citep{fischl2012freesurfer} and parcellated into 35 cortical regions of interest (ROIs) using Desikan-Killiany cortical atlas \citep{desikan2006automated}. The corresponding connectivity strength between two ROIs is derived by computing the absolute difference in their average cortical attribute (e.g., thickness) as introduced in \citep{mahjoub2018brain,nebli2020gender}. Table~\ref{tab:datadistribution} shows the distribution of both datasets.

\begin{table}[ht]
\centering
\begin{tabular}{ccccc}
\hline
\multirow{2}{*}{datasets} & \multicolumn{2}{c}{M/F} & \multicolumn{2}{c}{AD/LCMI} \\ \cline{2-5} 
                          & M          & F          & AD          & LMCI          \\ \hline
Number of subjects        & 615           & 781             & 70          & 64           \\
Right hemisphere (RH)                        & 308           & 391             & 35          & 32           \\
Left hemisphere (LH)                        & 308           & 391             & 35          & 32           \\
mean ± std. age      & 21.6$\pm{0.9} $       & 21.6 $\pm{0.8} $      &           &  \\ \hline
number of views           & 4             & 4               &4            & 4             \\ \hline
\end{tabular}
\\
\caption{\label{tab:datadistribution} Data distribution for M/F and AD/LCMI datasets, respectively. Each brain multigraph is represented by 4 connectivity views derived from cortical morphological measurements including maximum principal curvature, mean cortical thickness, mean sulcal depth, and average curvature.}
\end{table}

\subsection{Evaluation strategies}

We evaluate the performance of the single-view fusion methods set including SNF \citep{wang2014similarity}, NAG-FS \citep{mhiri2020joint}, and SM-netFusion \citep{mhiri2020supervised}, and the multigraph integration methods set including DGN \citep{gurbuz2020deep}, cMGI-Net \citep{demir2020clustering}, netNorm \citep{dhifallah2020estimation}, MVCF-Net \citep{chaari2020estimation}, and SCA \citep{dhifallah2019clustering}, separately. For CBT generation, we use as input dataset single-view networks for single-view CBT estimation and multi-view networks composed of 4 cortical morphological brain graphs for multigraph integration methods into a holistic CBT. To ensure the reproducibility and the generalizability of our evaluation results, we split each dataset into training and testing subsets using 5-fold cross-validation. We use the training subset to train the aforementioned 8 different models and to generate CBTs for both hemispheres (LH and RH) of 4 populations namely; AD, LMCI, M, and F. Next, we showcase each fusion method with four different evaluation tests on the left-out testing subset: (1) centeredness, (2) biomarker discovery of most discriminative connections between two groups, (3) graph global-level similarity to the original dataset, (4) graph node-wise similarity, and (5) graph distance-based similarity.

\subsection{Parameter settings}

We set all the hyperparameters for each graph fusion method using a grid search method. For SNF \citep{wang2014similarity}, we empirically set the number of nearest neighbors to $K= 20$ and the number of iterations $N_t = 20$ for convergence. For NAG-FS \citep{mhiri2020joint} method, we set the number of cluster to $N_c = 3$ for multiple kernels learning parameters. We also set the number of iterations to $N_t = 20$ for SNF parameters. Concerning the number of nearest neighbors for both SNF and multiple kernels learning, we opted for setting them to $K = 20$ which produced the best performance. For SM-netFusion \citep{mhiri2020supervised} parameters, we tested it using $N_c = 3$ clusters given the best result. For the cross-diffusion process parameters, we also set the number of iterations to $N_t = 20$ for convergence as demonstrated in \citep{wang2014similarity}. We fixed the number of closest neighbors $K = 20$. For SCA \citep{dhifallah2019clustering} parameters, we set the number of iterations to $N_t = 20$ as it guarantees SNF convergence \citep{wang2014similarity}. We set the number of nearest neighbors to $K = 20$ and for the clustering, we used $N_c = 3$ clusters giving the best results. For MVCF-Net \citep{chaari2020estimation} parameters, we set the number of clusters to $N_c = 3$ achieving the best results and the number of nearest neighbors to $K = 5$ for K-Nearest Neighbor(KNN) algorithm as recommended in \citep{chaari2020estimation}. For netNorm \citep{dhifallah2020estimation} parameters, we also set the the number of iterations used in SNF to $N_t = 20$ to guarantee its convergence as recommended in \citep{wang2014similarity}. We empirically set the number of nearest neighbors to $K = 20$. For cMGINet \citep{demir2020clustering} parameters, we set the number of clusters to $N_c = 3$ and the number of kernels to 5 for the multiple kernel learning. We trained the cMGINet model using 300 epochs with hyperparameters $\lambda = 0.3$ for scaling the subject-specific integration loss and the number of channels $n_c = 2$. For DGN \citep{gurbuz2020deep} parameters, we empirically set the hyperparameters to 3 edge-conditioned convolutional neural network layers with an edge-conditioned filter learner neural network. These layers are separated by ReLU activation function and output embeddings with 36, 24, and 5 dimensions for each ROI in the multigraph brain networks, respectively. The DGN is trained using gradient descent with Adam optimizer and a learning rate of 0.0005. The number of random samples in the subject normalization loss function is fixed to 10.

\subsection{CBT centeredness test}

\begin{figure}[H]
\centering
\includegraphics[width=1\linewidth]{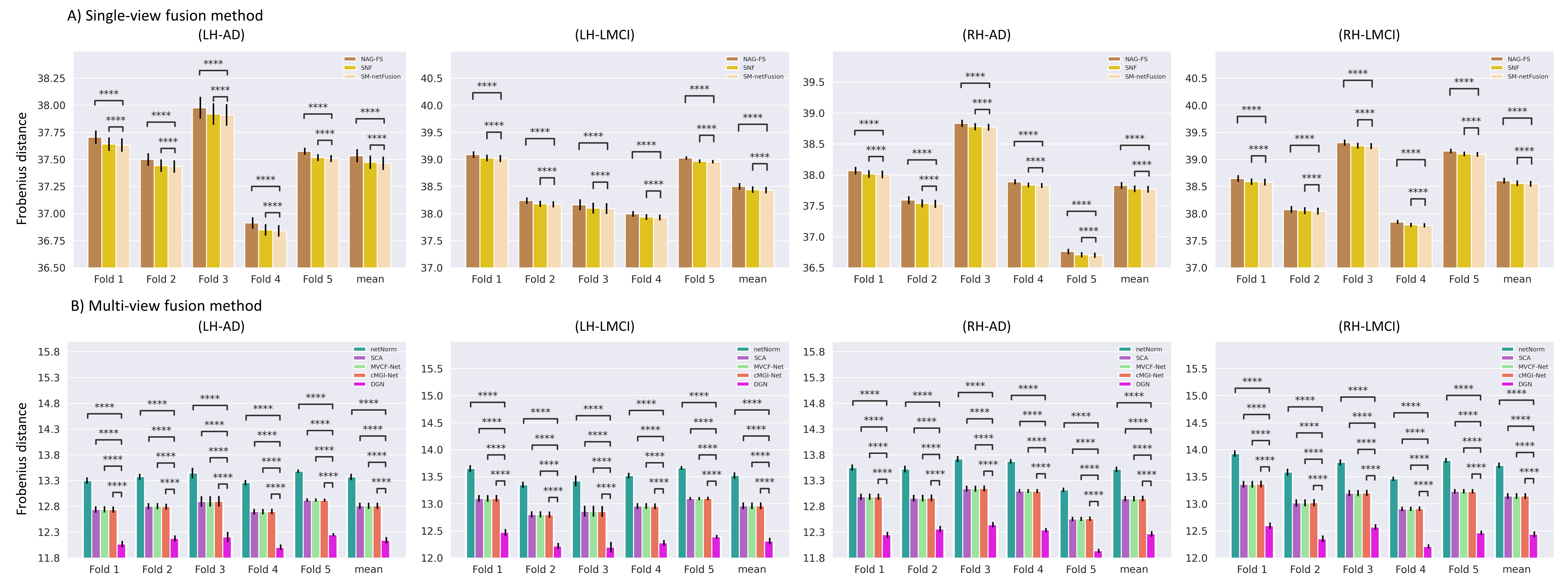}
\caption{Centeredness comparison of connectional templates generated by \textbf{A)} single-view integration methods including network atlas-guided feature selection (NAG-FS) \citep{mhiri2020joint}, similarity network fusion (SNF) \citep{wang2014similarity}, and supervised multi-topology network cross-diffusion (SM-netFusion) \citep{mhiri2020supervised}; and \textbf{B)} multi-graph fusion methods including multi-view networks normalizer (netNorm) \citep{dhifallah2020estimation}, cluster-based network fusion (SCA) \citep{dhifallah2019clustering}, multi-view clustering and fusion (MVCF-Net) \citep{chaari2020estimation}, cluster-based multi-graph integrator networks (cMGI-Net) \citep{demir2020clustering}, and deep graph normalizer (DGN) \citep{gurbuz2020deep}. Charts illustrate the mean Frobenius distance between the connectional templates learned from the training sets and networks of the samples in the testing set using a 5-fold cross-validation strategy. We reported the average distance for each cross-validation fold as well as the average across folds (``Mean'' bars on the right). For multi-graph fusion methods comparison, DGN achieved the lowest mean Frobenius distance to the population multi-view networks with a high statistical significance demonstrated by a two-tailed paired t-test (all $p < 0.0001$) for DGN-SCA, DGN-netNorm, DGN-MVCFNet, and DGN-cMGI-Net pairs for AD-LH, AD-RH, LMCI-LH and LMCI-RH groups. LH: left hemisphere. RH: right hemisphere. AD: Alzheimer's disease. LMCI: Late Mild Cognitive Impairment. As for single-view fusion methods comparison, SM-netFusion significantly achieved the lowest mean Frobenius distance to the population single-view networks (all $p < 0.0001$) using two-tailed paired t-test for SM-netFusion-NAGFS, and SM-netFusion-SNF pairs for AD-LH, AD-RH, LMCI-LH and LMCI-RH datasets.}
\label{fig:FrobeniusDistance}
\end{figure}

We evaluate the centeredness of the estimated CBT by measuring its Frobenius distance to each tensor view of each subject in the unseen test population. According to the results in \textbf{Fig.\ref{fig:FrobeniusDistance}} and \textbf{Supplementary information Fig. 1}, DGN \citep{gurbuz2020deep} largely outperforms all benchmark multigraph integration methods by achieving the minimum average Frobenius distance for all evaluation datasets (AD/LMCI and M/F datasets), for subpopulations (5-fold) and the mean over the folds in both hemispheres (RH and LH). For single-view fusion methods comparison, SM-netFusion \citep{mhiri2020supervised} slightly outperforms SNF and NAG-FS by achieving a lower Frobenius distance value than other methods. We note that DGN and SM-netFusion significantly outperform other methods across all left-out folds and evaluation datasets using  two-tailed paired t-test with a reproducible $p < 0.0001$ (\textbf{Fig.\ref{fig:FrobeniusDistance}} and \textbf{Supplementary information Fig. 1}). These results can be explained by the fact that, unlike other single-view and multigraph fusion methods, DGN integrates a randomized weighted loss function that acts as a regularizer to minimize the distance between the population of multi-graph brain networks and the estimated CBT, thereby enforcing its centeredness. DGN also refines the estimated CBT using a post-training process based on the element-wise median of all training CBTs to select the most centered connections for the final CBT generation. Most importantly, it is trained in an end-to-end manner.  

\subsection{CBT discriminative biomarker reproducibility test}

In addition to being well-centered, we demonstrate that DGN generates a well-discriminative CBT able to easily spot both gender-distinctive brain regions and AD-LMCI-distinctive brain regions. This can be explained by the fact that DGN captures the most discriminative brain connectivities of a population of multigraph networks, acting as connectional biomarkers. Particularly, we first spot the top $k$ ($k = 10, 15, 20, 25$) most discriminative brain regions distinguishing between two populations (e.g., AD/LMCI) for each cortical hemisphere using the estimated CBTs representing each class. Next, to evaluate the reproducibility of CBT-based discriminative ROIs, we train a support vector machine (SVM) to learn how to classify two populations coupled with Multiple Kernel Learning (MKL) to learn a weight vector that scores the discriminativeness of each feature (i.e., ROI). Next, we compute the overlap between the most discriminative ROIs identified using inter-class CBT difference and those using MKL. Table \ref{tab:overlap} displays the overlap in $\%$ between the top 10, 15, 20, and 25 discriminative ROIs identified using (i) MKL and (ii) the absolute difference between two estimated CBTs for each class generated by all single-view and multigraph fusion methods, respectively, using both AD/LMCI datasets and M/F connectomic datasets. We demonstrate that DGN method reaches the highest overlap percentage with respect to other multigraph and single-view fusion methods between AD/LMCI datasets by achieving a boost of 14-32$\%$ and 20-46$\%$ in biomarker reproducibility against other methods in the left and the right hemispheres, respectively. Furthermore, DGN ranked first in reproducibility where it achieves 12-30$\%$ and 12-40$\%$ boost in identifying the most discriminative brain regions between genders against other methods in the left and right hemispheres, respectively, using M/F datasets (Table \ref{tab:overlap}). The displayed results represent respectively the minimum and the maximum differences between the reproducibility rates of DGN and each of the other methods over the top $k$ (10, 15, 20, and 20) discriminative ROIs distinguishing between AD and LMCI populations, and between male and female populations.

\begin{table}[ht]
\centering
\resizebox{\textwidth}{!}{%
\begin{tabular}{|c|c|c|c|c|c|c|c|c|c|c|c|c|c|c|c|c|c|}
\hline
\multicolumn{2}{|c|}{Top k (\%) discriminative ROIs}                                                                                       & \multicolumn{4}{c|}{Matching rate (10 \% )}                                                                                                           & \multicolumn{4}{c|}{Matching rate (15 \% )}                                                                                                               & \multicolumn{4}{c|}{Matching rate (20 \% )}                                                                                                               & \multicolumn{4}{c|}{Matching rate (25 \% )}                                                                                                               \\ \hline
\multicolumn{2}{|c|}{datasets}                                                                                                             & \multicolumn{2}{c|}{AD-LMCI}                                              & \multicolumn{2}{c|}{GSP}                                                  & \multicolumn{2}{c|}{AD-LMCI}                                                & \multicolumn{2}{c|}{GSP}                                                    & \multicolumn{2}{c|}{AD-LMCI}                                                & \multicolumn{2}{c|}{GSP}                                                    & \multicolumn{2}{c|}{AD-LMCI}                                                & \multicolumn{2}{c|}{GSP}                                                    \\ \hline
\multicolumn{2}{|c|}{Hemispheres}                                                                                                          & LH                                  & RH                                  & LH                                  & RH                                  & LH                                   & RH                                   & LH                                   & RH                                   & LH                                   & RH                                   & LH                                   & RH                                   & LH                                   & RH                                   & LH                                   & RH                                   \\ \hline
                                                                                            & SNF\citep{wang2014similarity}                                          & 0.3                                 & 0.1                                 & 0.1                                 & 0.1                                 & 0.2                                  & 0.13                                 & 0.33                                 & 0.4                                  & 0.5                                  & 0.5                                  & 0.5                                  & 0.5                                  & 0.68                                 & 0.68                                 & 0.68                                 & 0.64                                 \\ \cline{2-18} 
                                                                                            & {\color[HTML]{7D29EB} \textbf{ SM-netFusion \citep{mhiri2020supervised}}} & {\color[HTML]{7D29EB} \textbf{0.5}} & {\color[HTML]{7D29EB} \textbf{0.5}} & {\color[HTML]{7D29EB} \textbf{0.1}} & {\color[HTML]{7D29EB} \textbf{0.1}} & {\color[HTML]{7D29EB} \textbf{0.47}} & {\color[HTML]{7D29EB} \textbf{0.53}} & {\color[HTML]{7D29EB} \textbf{0.33}} & {\color[HTML]{7D29EB} \textbf{0.46}} & {\color[HTML]{7D29EB} \textbf{0.55}} & {\color[HTML]{7D29EB} \textbf{0.55}} & {\color[HTML]{7D29EB} \textbf{0.55}} & {\color[HTML]{7D29EB} \textbf{0.55}} & {\color[HTML]{7D29EB} \textbf{0.72}} & {\color[HTML]{7D29EB} \textbf{0.72}} & {\color[HTML]{7D29EB} \textbf{0.72}} & {\color[HTML]{7D29EB} \textbf{0.72}} \\ \cline{2-18} 
\multirow{-3}{*}{\begin{tabular}[c]{@{}c@{}}Single-\\ view\\ fusion\\ methods\end{tabular}} & NAG-FS \citep{mhiri2020joint}                                     & 0.2                                 & 0.2                                 & 0                                   & 0.2                                 & 0.27                                 & 0.27                                 & 0.27                                 & 0.33                                 & 0.5                                  & 0.5                                  & 0.46                                 & 0.5                                  & 0.68                                 & 0.68                                 & 0.64                                 & 0.64                                 \\ \hline
                                                                                            & SCA\citep{dhifallah2019clustering}                                         & 0.3                                 & 0.3                                 & 0.2                                 & 0.4                                 & 0.4                                  & 0.53                                 & 0.27                                 & 0.33                                 & 0.45                                 & 0.65                                 & 0.4                                  & 0.5                                  & 0.64                                 & 0.68                                 & 0.68                                 & 0.68                                 \\ \cline{2-18} 
                                                                                            & MVCF-Net\citep{chaari2020estimation}                                     & 0.46                                & 0.46                                & 0.2                                 & 0.4                                 & 0.47                                 & 0.53                                 & 0.27                                 & 0.33                                 & 0.45                                 & 0.55                                 & 0.4                                  & 0.5                                  & 0.64                                 & 0.68                                 & 0.68                                 & 0.68                                 \\ \cline{2-18} 
                                                                                            & netNorm\citep{dhifallah2020estimation}                                      & 0.3                                 & 0.4                                 & 0.2                                 & 0.4                                 & 0.4                                  & 0.53                                 & 0.33                                 & 0.4                                  & 0.55                                 & 0.6                                  & 0.5                                  & 0.5                                  & 0.72                                 & 0.64                                 & 0.7                                  & 0.64                                 \\ \cline{2-18} 
                                                                                            & cMGI-Net\citep{demir2020clustering}                                      & 0.2                                 & 0.4                                 & 0                                   & 0.1                                 & 0.4                                  & 0.47                                 & 0.27                                 & 0.33                                 & 0.55                                 & 0.65                                 & 0.4                                  & 0.45                                 & 0.72                                 & 0.72                                 & 0.64                                 & 0.68                                 \\ \cline{2-18} 
\multirow{-5}{*}{\begin{tabular}[c]{@{}c@{}}Multigraph\\ fusion\\ methods\end{tabular}}     & {\color[HTML]{CE25E6} \textbf{DGN\citep{gurbuz2020deep}}}          & {\color[HTML]{CE25E6} \textbf{0.5}} & {\color[HTML]{CE25E6} \textbf{0.5}} & {\color[HTML]{CE25E6} \textbf{0.3}} & {\color[HTML]{CE25E6} \textbf{0.5}} & {\color[HTML]{CE25E6} \textbf{0.53}} & {\color[HTML]{CE25E6} \textbf{0.6}}  & {\color[HTML]{CE25E6} \textbf{0.53}} & {\color[HTML]{CE25E6} \textbf{0.53}} & {\color[HTML]{CE25E6} \textbf{0.6}}  & {\color[HTML]{CE25E6} \textbf{0.7}}  & {\color[HTML]{CE25E6} \textbf{0.6}}  & {\color[HTML]{CE25E6} \textbf{0.6}}  & {\color[HTML]{CE25E6} \textbf{0.76}} & {\color[HTML]{CE25E6} \textbf{0.84}} & {\color[HTML]{CE25E6} \textbf{0.76}} & {\color[HTML]{CE25E6} \textbf{0.76}} \\ \hline
\end{tabular}%
}
\\
\caption{\label{tab:overlap} Matching rate in \% between the top $k$ (10, 15, 20,and 20) discriminative ROIs distinguishing between AD and LMCI poulations, and between male and female populations identified by (i) MKL \citep{varma2009more} and CBT-based single-view fusion methods and (ii) MKL \citep{varma2009more} and CBT-based multigraph fusion methods for the right and left hemispheres (RH and LH).}
\end{table}

\begin{sidewaysfigure}
\centering
\includegraphics[width=19.5cm]{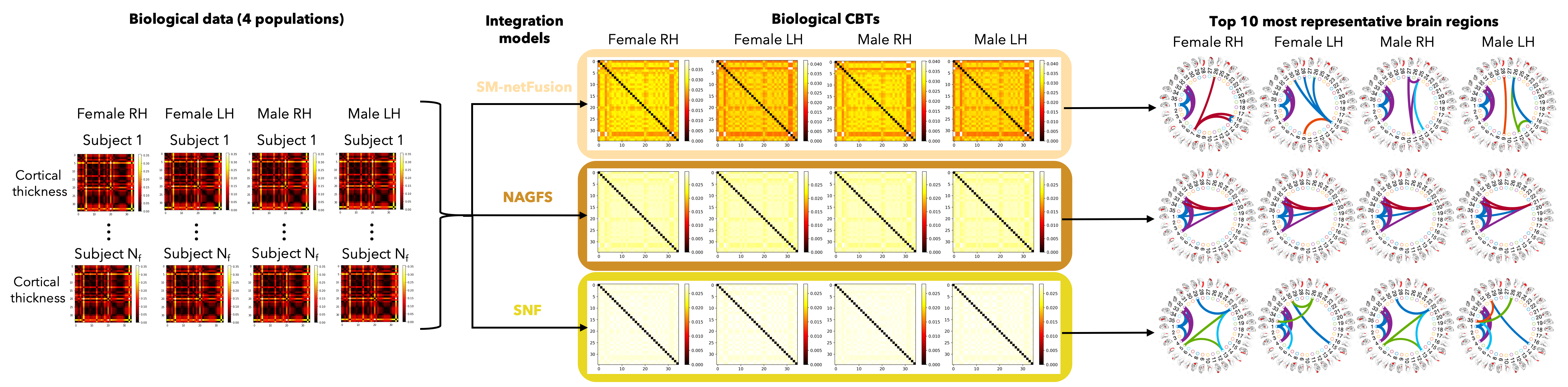}
\caption{Integrating single-view brain graph data integration and extracting the top 10 most representative brain connectivities. We display the learned CBTs by single-view integration methods including network atlas-guided feature selection (NAG-FS) \citep{mhiri2020joint}, similarity network fusion (SNF) \citep{wang2014similarity}, and supervised multi-topology network cross-diffusion (SM-netFusion) \citep{mhiri2020supervised} for the left hemisphere (LH) and the right hemisphere (RH) of the female (F) and male (M) populations. It is apparent that the templates generated by SM-netFusion encapsulate topological patterns which commonly exist in all subjects of each group (eg., LH-F, LH-M, RH-F, RH-M) represented by single-view brain networks. As for NAGFS and SNF, they capture only a few local motifs across subjects in a population. The circular graphs display the top 10 most representative cortical morphological connections between brain regions having the highest weights, of male and female brain networks groups, respectively, for both right and left hemispheres.}
\label{fig:biologicalCBTSingle-view} 
\end{sidewaysfigure}

\begin{sidewaysfigure}
\centering
\includegraphics[width=19.5cm]{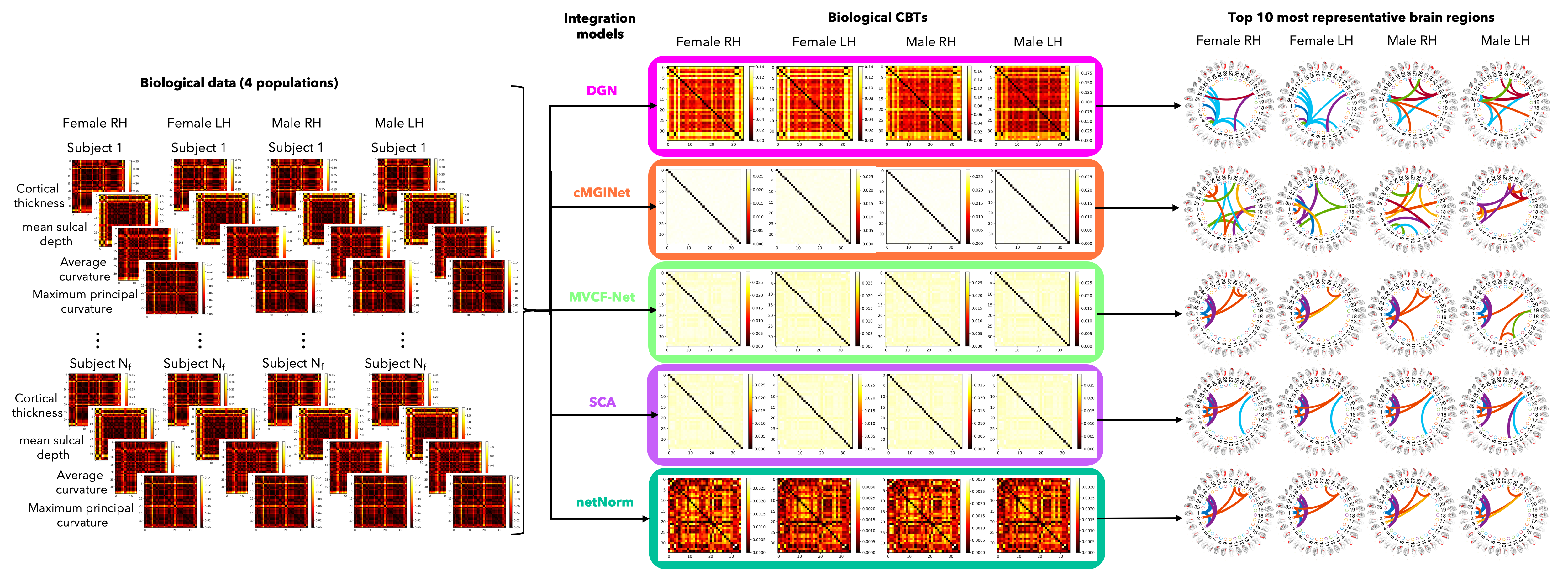}
\caption{Integrating multi-view brain multigraph data integration and extracting the top 10 most representative brain connectivities. We display the CBTs estimated by multigraph integration methods including multi-view networks normalizer (netNorm) \citep{dhifallah2020estimation}, cluster-based network fusion (SCA) \citep{dhifallah2019clustering}, multi-view clustering and fusion (MVCF-Net) \citep{chaari2020estimation}, cluster-based multi-graph integrator networks (cMGI-Net) \citep{demir2020clustering}, and deep graph normalizer (DGN) \citep{gurbuz2020deep} for the left hemisphere (LH) and the  right hemisphere (RH) of female (F) and male (M) groups. It is apparent that the templates generated by DGN encapsulate topological patterns which commonly exist in all views of all subjects of each group (eg., LH-F, LH-M, RH-F, RH-M). Circular graphs display the top 10 most representative cortical morphological connections between brain regions having the highest weights, of male and female brain networks groups, respectively, for both right and left hemispheres, allowing us to easily identify cross-population discriminative brain connectivities.}
\label{fig:biologicalCBTmulti-view}
\end{sidewaysfigure}

\subsection{CBT node-wise similarity comparison}

\subsubsection{Hubness behaviour test}

\begin{figure}[H]
\centering
\includegraphics[width=0.98\linewidth]{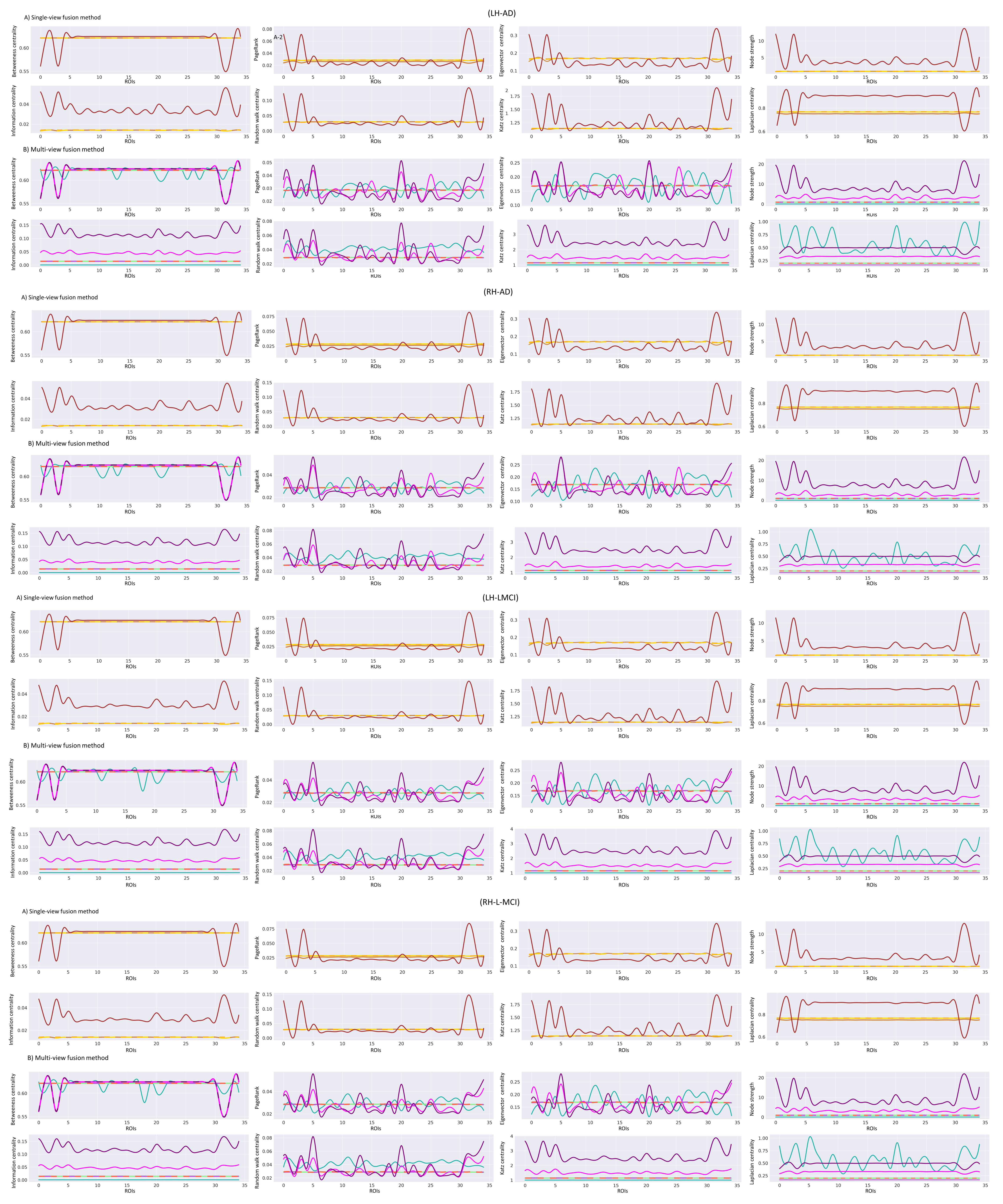}
\caption{Comparison of the average topological distributions across 5-fold cross-validation of PageRank \citep{xing2004weighted}, Katz centrality \citep{katz1953new}, node strength \citep{barrat2004architecture}, random-walk centrality \citep{newman2005measure}, information centrality \citep{stephenson1989rethinking}, Laplacian centrality \citep{qi2012laplacian}, eigenvector centrality \citep{newman2008mathematics}, and betweeness centrality \citep{brandes2001faster} of templates generated by \textbf{A)} SNF\citep{wang2014similarity}, NAG-FS \citep{mhiri2020joint}, and SM-netFusion \citep{mhiri2020supervised} against the ground truth distribution for a population of single-view network; and \textbf{B)} netNorm \citep{dhifallah2020estimation}, SCA \citep{dhifallah2019clustering}, MVCF-Net \citep{chaari2020estimation}, cMGI-Net\citep{demir2020clustering}, and DGN\citep{gurbuz2020deep} against the ground truth distribution for a population of multi-view network for AD and LMCI datasets in the left and right hemispheres.}
\label{fig:hubness}
\end{figure}

Furthermore, we compare the performance of the graph fusion methods by evaluating the similarity of their generated CBTs at the node-wise scale. One way to do so is to quantify the topological properties of the CBTs using the following centrality measures: node strength, betweenness centrality, random-walk betweenness centrality, eigenvector centrality, weighted PageRank, Katz centrality, information centrality, and Laplacian centrality. Specifically, we evaluate the topological properties of the learned CBTs by comparing the likelihood of distribution of each topological measure between the ground truth of the multigraph brain population and the learned CBTs. We calculate the ground truth by simply averaging the distribution of topological measures of each network view of each testing subject. We display the average across five folds for each centrality metric in the form of distribution graphs for AD (LH), LMCI (LH), AD (RH), and LMCI (LH) populations (\textbf{Fig~\ref{fig:hubness}}) and for M (LH), F (LH), M (RH), and F (LH) populations (\textbf{Supplementary information Fig. 2}). 

\begin{figure}[H]
\centering
\includegraphics[width=1\linewidth]{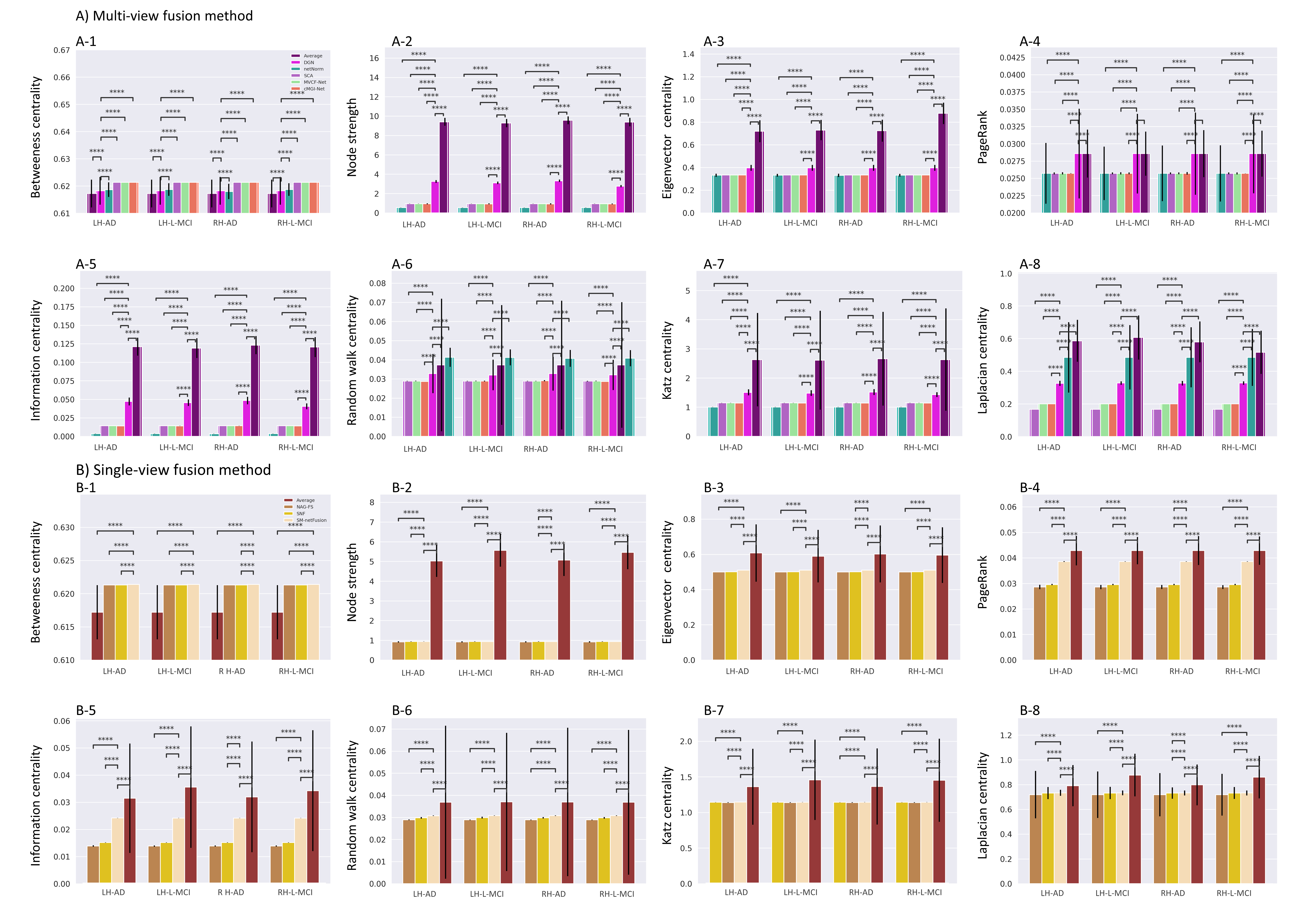}
\caption{This chart displays the average topological distributions of PageRank \citep{xing2004weighted}, Katz centrality \citep{katz1953new}, node strength \citep{barrat2004architecture}, random-walk centrality \citep{newman2005measure}, information centrality\citep{stephenson1989rethinking}, Laplacian centrality \citep{qi2012laplacian}, eigenvector centrality \citep{newman2008mathematics}, and betweeness centrality \citep{brandes2001faster} measures across the nodes (ROIs) of the learned templates generated by \textbf{A)} SNF \citep{wang2014similarity}, NAG-FS \citep{mhiri2020joint}, and SM-netFusion \citep{mhiri2020supervised} against the ground truth distribution of a population of single-view networks; and \textbf{B)} netNorm \citep{dhifallah2020estimation}, SCA \citep{dhifallah2019clustering}, MVCF-Net \citep{chaari2020estimation}, cMGI-Net\citep{demir2020clustering}, and DGN\citep{gurbuz2020deep} against the ground truth distribution of a population of multi-view network for the AD and LMCI datasets in the left and right hemispheres. For multi-graph fusion methods comparison, DGN achieved the highest average distribution comparing to the average distribution of other multigraph fusion methods with a high statistical significance demonstrated by a two-tailed paired t-test (all $p < 0.0001$) for DGN-SCA, DGN-netNorm, DGN-MVCFNet, and DGN-cMGI-Net pairs for AD (LH), AD (RH), LMCI (LH) and LMCI (RH) groups, except for the random-walk centrality measures. For single-view fusion methods comparison,SM-netFusion significantly achieved the maximum average distribution comparing to SNF and NAG-FS for all centrality measures except the betweeness centrality and node strength for AD (LH), AD (RH), LMCI (LH) and LMCI (RH) datasets.}
\label{fig:averagehubness}
\end{figure}

As shown in \textbf{Fig~\ref{fig:hubness}} and \textbf{Supplementary information Fig. 2}, the connectional brain template generated by DGN shows a striking similarity with the ground truth data in topological properties while other multigraph integration methods and all single-view fusion methods fail to preserve the multi-view and single-view connectomic data topology, respectively. This can be explained by the fact that DGN has the ability to capture much more complex topological patterns in comparison with other fusion architectures and in a fully generic manner \textbf{Fig~\ref{fig:biologicalCBTmulti-view}}. Specifically, DGN trains a GDL-based learning process for brain connection weights by blending a sequence of hidden nodes embeddings with the integrated connectivities while capturing complex patterns and non-linear variation across individuals. For easy interpretation and better visualization of the results, we compute the average of each centrality measure distribution across the ROIs, so that each distribution is represented by a single value. The results in \textbf{Fig~\ref{fig:averagehubness}} (for AD/LMCI dataset) and \textbf{Supplementary information Fig. 3} (for M/F dataset) confirm that DGN is the most topology-preserving method in a population of multi-view networks by closely nearing the average distribution of the ground truth multi-view brain networks. 

For single-view fusion methods, SM-netFusion achieved the closest average hubness distribution to that of ground truth. This can be explained by the fact that SM-netFusion encapsulates topological patterns which commonly exist in all single-view brain networks comparatively with other single-view integration methods. As for NAGFS and SNF methods, they capture only a few local motifs across subjects in a population \textbf{Fig~\ref{fig:biologicalCBTSingle-view}}. Using a weighted mixture of multi-topological measures to boost the non-linear fusion of single-view networks, SM-netFusion integrates complementary information of single-view brain networks across all individuals in the population. For each biological CBT learned by single-view or multigraph fusion methods, we identified the top 10 weighted cortical morphological connectivities fingerprinting the input brain graph population. The selected CBT edges represent the most representative and holistic brain connectivities following the integration process (\textbf{Fig}~\ref{fig:biologicalCBTSingle-view}-\ref{fig:biologicalCBTmulti-view}).

\begin{figure}[H]
\centering
\includegraphics[width=1\linewidth]{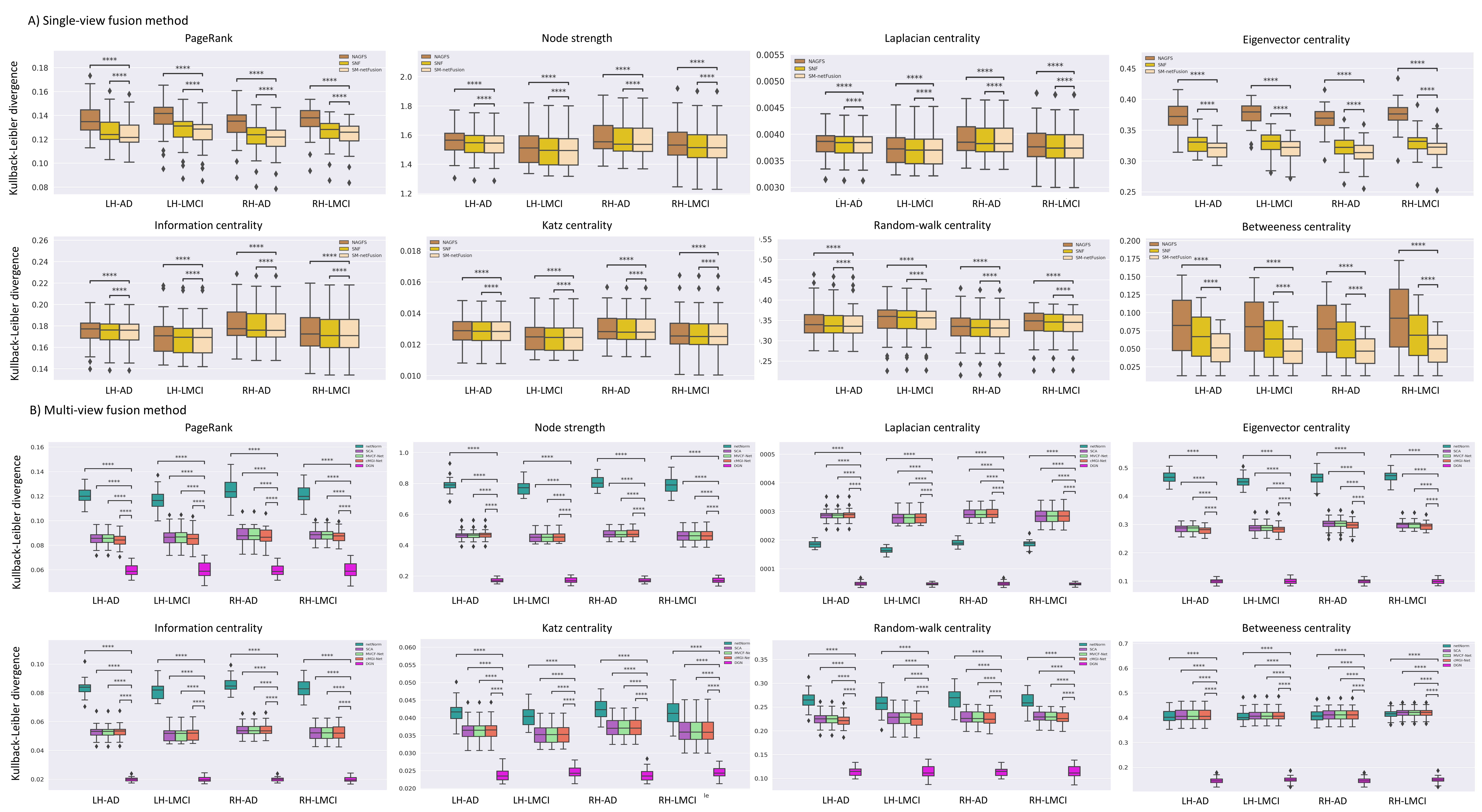}
\caption{ Average of Kullback-Liebler divergence distribution across 5-fold cross validation between the ground truth distribution and the average topological distributions of the learned connectional templates generated by \textbf{A)} single-view fusion methods (SNF \citep{wang2014similarity}, NAG-FS \citep{mhiri2020joint},and SM-netFusion \citep{mhiri2020supervised}); and \textbf{B)} multigraph fusion methods (SCA \citep{dhifallah2019clustering}, netNorm \citep{dhifallah2020estimation}, MVCF-Net \citep{chaari2020estimation}, cMGI-Net\citep{demir2020clustering}, and DGN\citep{gurbuz2020deep}). The topological measures include PageRank \citep{xing2004weighted}, Katz centrality \citep{katz1953new}, node strength \citep{barrat2004architecture}, random-walk centrality \citep{newman2005measure}, information centrality \citep{stephenson1989rethinking}, Laplacian centrality \citep{qi2012laplacian}, eigenvector centrality \citep{newman2008mathematics}, and betweeness centrality \citep{brandes2001faster}. Charts illustrate that for multi-graph fusion methods comparison, DGN achieved the minimum mean KL-divergence distribution and the narrowest dispersion range with a high statistical significance demonstrated by a two-tailed paired t-test (all $p < 0.0001$) for DGN-SCA, DGN-netNorm, DGN-MVCFNet, and DGN-cMGI-Net pairs for AD-LH, AD-RH, LMCI-LH and LMCI-RH groups. For single-view fusion methods comparison, SM-netFusion significantly achieved the lowest mean KL-divergence distribution to the population single-view networks (all $p < 0.0001$) using two-tailed paired t-test for SM-netFusion-NAGFS, and SM-netFusion-SNF pairs for AD-LH, AD-RH, LMCI-LH and LMCI-RH datasets. [Box plot legend: median (midline), box (25th and 75th percentiles), and whiskers (extrema).]}
\label{fig:kullbackDivergence}
\end{figure}

Second, we extend the hubness comparison between the CBTs by adding the Kullback-Liebler (KL) divergence which measures the dissimilarity between two given graphs by quantifying the information change between them. The main idea was to compute for each centrality metric the KL-divergence of (i) the ground truth distribution and (ii) each of the distributions derived from the connectional brain templates, learned within a 5-fold cross-validation strategy. Note that we normalized each distribution using the sum over all nodes to get a valid discrete probability distribution. Next, we reported the KL-divergence distribution resulting from all possible pair combinations of sub-populations (folds). We evaluated the performance of the graph fusion methods based on the lowest value of the average KL-divergence distribution and its standard deviation over all combinations of 5 sub-populations. A small divergence signifies similar distributions. Comparing to other multigraph fusion methods \textbf{Fig~\ref{fig:kullbackDivergence}} and \textbf{Supplementary information Fig. 4} shows that DGN significantly outperforms other multigraph fusion methods on both evaluation datasets using the left and right hemispheres and across all topological measures (two-tailed paired t-test $p < 0.0001$) by achieving the minimum scores in both mean KL-divergence distribution and its dispersion for AD, LMCI, M, and F datasets. We demonstrated that DGN generates the most similar centrality measure distribution to the ground truth by preserving the complex patterns in a population of multi-view networks during the data integration process to generate more holistic and integral connectional templates. For the single-view fusion methods comparison, \textbf{Fig~\ref{fig:kullbackDivergence}} and \textbf{Supplementary information Fig. 4} showed that SM-netFusion outperforms SNF and NAG-FS by achieving the minimum mean KL-divergence distribution in AD/LMCI dataset and M/F dataset, respectively, for both hemispheres.

\begin{figure}[H]
\centering
\includegraphics[width=0.96\linewidth]{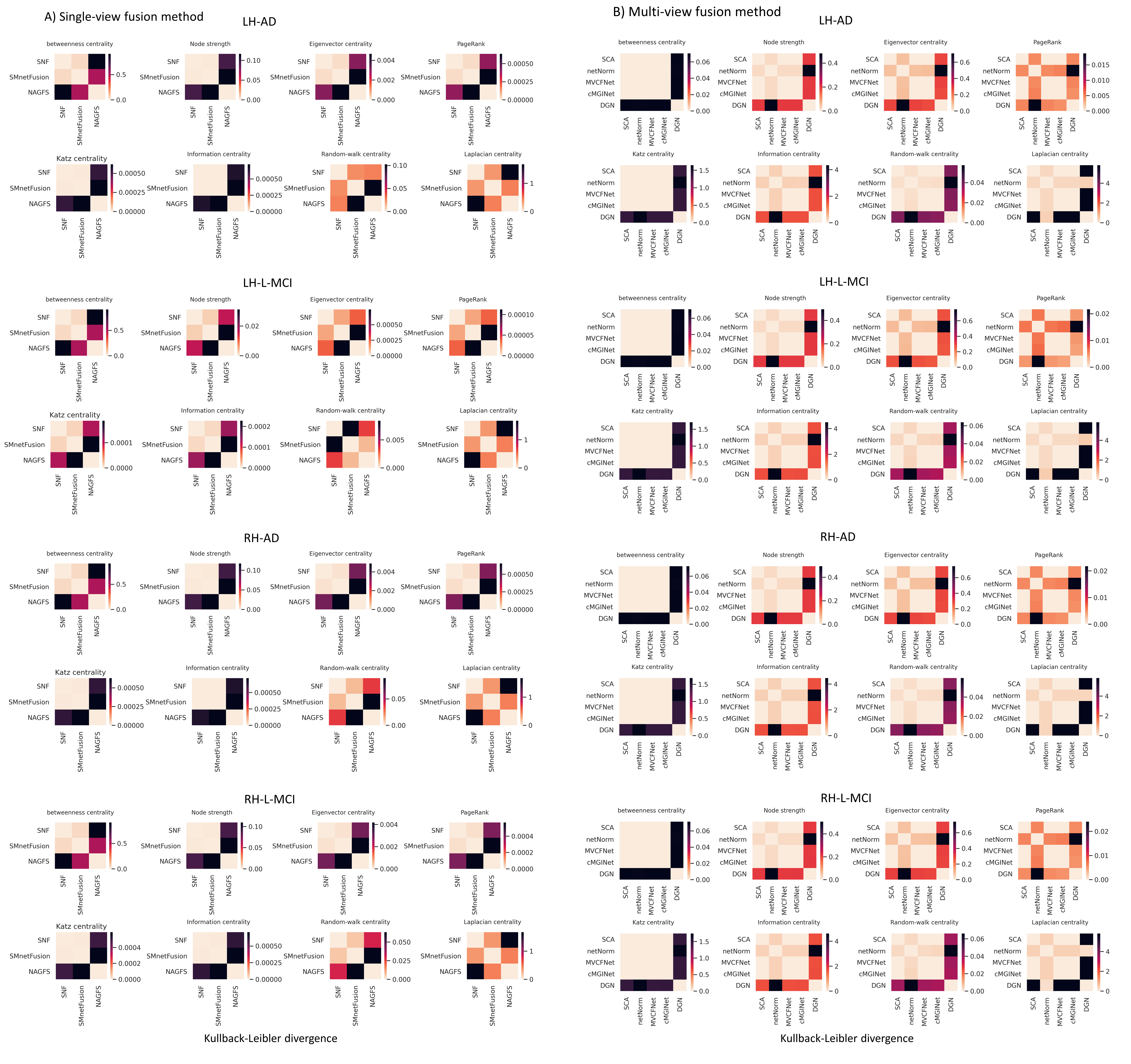}
\caption{Dissimilarity between the possible pairs combination of \textbf{A)} single-view and \textbf{B)} multigraph fusion methods using KL-divergence for the AD and LMCI in the left and the right hemispheres.}
\label{fig:pairwiseKullbackDivergence}
\end{figure}

Lastly, we compute a pairwise comparison between all combinations of the single-view and multigraph fusion methods, separately, using KL divergence. Specifically, for each centrality measure (CM), we average the CBT-based CM distribution across 5-folds sub-populations, then we compute the KL-divergence between a pair of normalized CBT-based CM distributions derived from two selected methods. The resulting score reflects the topological dissimilarity between the pair of methods. \textbf{Fig~\ref{fig:pairwiseKullbackDivergence}} and \textbf{Supplementary information Fig. 5} display the KL-divergence between all possible pairs of single-view and multigraph fusion methods using AD/LMCI dataset and M/F dataset, respectively, for left and right hemispheres. Remarkably, DGN stands out again with the highest KL-divergence among the other multigraph fusion methods across all datasets and all centrality metrics, while SM-netFusion differs the most among other single-view fusion methods.

\subsubsection{Segregation behaviour test}

\begin{figure}[H]
\centering
\includegraphics[width=1\linewidth]{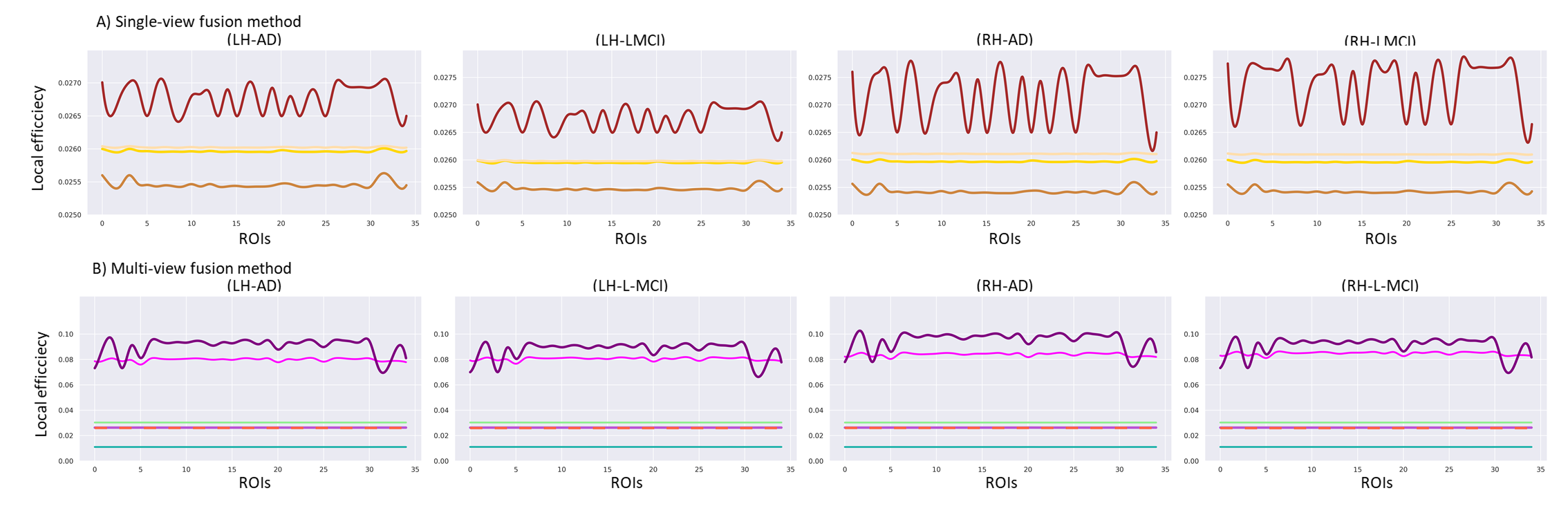}
\caption{ Region-wise local efficiency distribution of connectional brain templates generated by \textbf{A)} SNF \citep{wang2014similarity}, NAG-FS \citep{mhiri2020joint}, and SM-netFusion\citep{mhiri2020supervised} against the ground truth distribution for single-view fusion methods comparison; and \textbf{B)} netNorm \citep{dhifallah2020estimation}, SCA \citep{dhifallah2019clustering}, MVCF-Net \citep{chaari2020estimation}, cMGI-Net\citep{demir2020clustering}, and DGN\citep{gurbuz2020deep} against the ground truth for multi-graph integration methods using 5-fold cross-validation for the AD and LMCI populations in the left and right hemispheres. For multi-graph fusion methods comparison, DGN achieved the most similar distribution to the ground truth while SM-netFusion displayed the closest distribution.}
\label{fig:localefficiency}
\end{figure}

Another aspect to compare the performance of the graph fusion methods is by evaluating the segregation behavior of their generated CBTs. For that, we computed the local efficiency distribution across the ROIs of each connectional template (nodes graph) using 5-fold cross-validation. Next, we report the average local efficiency distribution over all sub-populations (testing folds) for AD, LMCI using left and right hemispheres. For a fair comparison, we further compute the ground truth distribution by averaging the local efficiency that is independently calculated for each view of each testing subject. \textbf{Fig~\ref{fig:localefficiency}} and \textbf{Supplementary information Fig. 6} shows that DGN has the most similar distribution with the ground truth for AD, LMCI, M, and F datasets using both hemispheres.  SM-netFusion displayed the closest distribution for single-view CBT estimation. For easy visualization of the results, we display the average local efficiency distribution over the ROIs for the ground truth brain networks and the learned CBT by different methods. Remarkably, \textbf{Fig~\ref{fig:integrationAndglobal-level}} and \textbf{Supplementary information Fig. 7} confirms that DGN achieves the highest average local efficiency over regions compared to other methods for AD (LH), AD (RH), LMCI (LH) and LMCI (RH) datasets. This can be explained by the fact that DGN aggregates the information passed by its neighbors while taking into consideration the multi-view attributes of its neighboring edges. This was done by integrating graph convolution layers which act as edge conditioned filter learners to learn deeper embeddings for each ROI. As result, the information is efficiently transferred to neighboring nodes while fusing the population of multigraph networks. For single-view fusion methods comparison, SM-netFusion slightly achieved the highest average local efficiency over regions comparing to SNF and NAG-FS for ASD/LMCI and M/F datasets (\textbf{Fig~\ref{fig:integrationAndglobal-level}} and {\textbf{Supplementary information Fig. 7}}).

\subsubsection{Integration behaviour test}

\begin{figure}[H]
\centering
\includegraphics[width=1\linewidth]{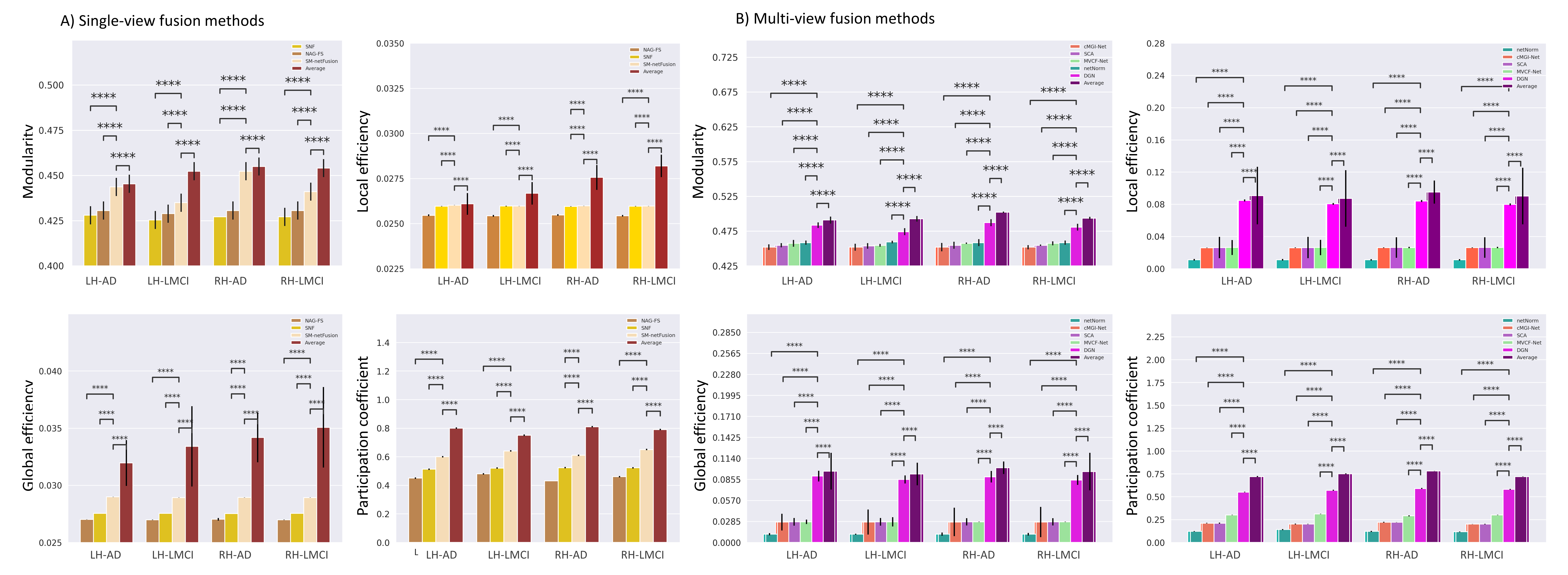}
\caption{Charts display the average local efficiency distribution across regions (ROIs), modularity, global efficiency, and participation coefficient of connectional brain templates estimated by \textbf{A)} single-view fusion methods (SNF \citep{wang2014similarity}, NAG-FS \citep{mhiri2020joint}, and SM-netFusion \citep{mhiri2020supervised}) and \textbf{B)} multigraph fusion methods (netNorm \citep{dhifallah2020estimation}, SCA \citep{dhifallah2019clustering}, MVCF-Net \citep{chaari2020estimation}, cMGI-Net\citep{demir2020clustering}, and DGN\citep{gurbuz2020deep}) against the ground truth for AD-LH, LMCI-LH, AD-RH, and LMCI-RH groups using 5-fold cross validation. Remarkably, DGN achieved the highest scores including average local efficiency distribution over regions, modularity, participation coefficient, and global efficiency comparing to other multigraph fusion methods with a high statistical significance demonstrated by a two-tailed paired t-test (all $p < 0.0001$) for DGN-SCA, DGN-netNorm, DGN-MVCFNet, and DGN-cMGI-Net pairs for AD-LH, AD-RH, LMCI-LH and LMCI-RH groups. While for the single-view integration methods comparison, SM-netFusion significantly outperformed SNF and NAG-FS for AD and LMCI datasets in both hemispheres (all $p < 0.0001$).}
\label{fig:integrationAndglobal-level}
\end{figure}

Furthermore, we compared the integration behavior between the estimated connectional templates using the coefficient participation metric, which quantifies the connection strength (node's edges) between communities (modules) in the graph for ASD/LMCI and M/F datasets \textbf{Fig~\ref{fig:integrationAndglobal-level}} and \textbf{Supplementary information Figure 7}. We compute the average participation coefficient across five folds for the CBTs generated by each graph fusion method and the ground truth network data, respectively. We acquire the ground truth by averaging the participation coefficient measures which are separately calculated for each testing sample and for each view network. We demonstrate that DGN achieves the maximum average participation coefficient compared to other multigraph fusion methods, and the closest score to the ground truth with a high statistical significance demonstrated by a two-tailed paired t-test (all $p < 0.0001$) for DGN-SCA, DGN-netNorm, DGN-MVCFNet, and DGN-cMGI-Net pairs for AD (LH), AD (RH), LMCI (LH) and LMCI (RH) groups (\textbf{Fig~\ref{fig:integrationAndglobal-level}} as well as for F (LH), F (RH), M (LH) and M (RH) populations (\textbf{Supplementary information Fig.7}. This can be explained by the fact that DGN learns the optimized integration of multigraph networks into a single representation population graph in an end-to-end manner using a GNN-based integrator while taking into account the data heterogeneity and complementary information across views within the same multigraph. For single-view fusion method comparison, SM-netFusion significantly achieves the maximum average participation coefficient compared to other single-view integration methods, and the closest score to the ground truth population with a high statistical significance demonstrated by a two-tailed paired t-test (all $p < 0.0001$) for SM-netFusion -SNF and SM-netFusion -NAG-FS pairs for AD (LH), AD (RH), LMCI (LH) and LMCI (RH) groups (\textbf{Fig~\ref{fig:integrationAndglobal-level}}) as well as for F (LH), F (RH), M (LH) and M (RH) populations (\textbf{Supplementary information Fig. 7}). This can be explained by the fact that SM-netFusion uses a weighted mixture of multi-topological measures to enhance the non-linear fusion of single-view networks, thus preserving the data topology in terms of integration behavior.

\subsection{CBT global-level similarity test}

As for the evaluation of the global-level similarity of the learned brain connectional templates, we include modularity and global efficiency measures. We compute the average modularity and the average global efficiency over the random sample partitions of the learned CBTs by different fusion methods. Next, we calculate the modularity and the global efficiency for each view of each testing sample and we averaged them to acquire the measures of the ground truth. We demonstrate once again that DGN significantly outperforms other multigraph fusion methods across evaluation datasets (AD, LMCI, M, and F) for both hemispheres (\textbf{Fig~\ref{fig:integrationAndglobal-level}} and \textbf{Supplementary information Fig. 7}, two-tailed paired t-test, $p < 0.0001$). This result is the outcome of the DGN learning process of fusing multigraph brain networks while preserving the strength of the connections in the entire graph structure across subjects, and thus preserving the brain graph communities. Specifically, DGN introduces a randomized weighted loss function (SLN) that optimizes connectivity weights of the generated CBT to ensure its representativeness in terms of community structures. For the single-view fusion methods comparison, SM-netFusion significantly achieved the maximum average modularity comparing to SNF and NAG-FS for AD, LMCI, M, and F in both left and right hemispheres (\textbf{Fig~\ref{fig:integrationAndglobal-level}} and \textbf{Supplementary information Fig. 7}), two-tailed paired t-test, $p < 0.0001$).

\subsection{CBT distance-based similarity test}	

\begin{figure}[H]
\centering
\includegraphics[width=1\linewidth]{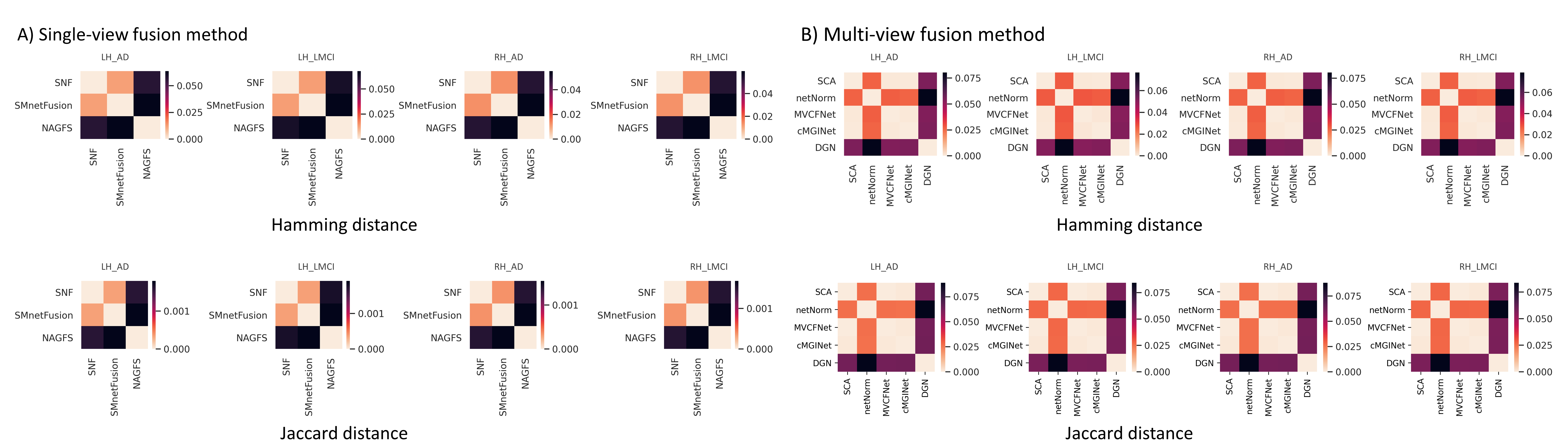}
\caption{Pairwise distance comparison of the learned connectional templates generated by \textbf{A)} single-view fusion methods (SNF \citep{wang2014similarity}, NAG-FS \citep{mhiri2020joint}, and SM-netFusion \citep{mhiri2020supervised}); and \textbf{B)} multigraph fusion methods (netNorm\citep{dhifallah2020estimation}, SCA \citep{dhifallah2019clustering}, MVCF-Net \citep{chaari2020estimation}, cMGI-Net \citep{demir2020clustering}, and DGN \citep{gurbuz2020deep}) using Hamming distance and Jaccard distance measures for AD and LMCI populations in the left and right hemispheres.}
\label{fig:Hamming-Jaccard}
\end{figure}

To evaluate the similarity between the estimated connectional templates by different methods, we compute both Hamming distance and Jaccard distance between all possible pairs of the learned connectional templates generated by single-view and multigraph fusion methods, separately, using AD, LMCI, M, and F for left and right hemispheres. Remarkably, the CBT learned by the DGN method stands out with the highest Hamming and Jaccard distances among CBTs generated by other multigraph fusion methods across all datasets (\textbf{Fig~\ref{fig:Hamming-Jaccard}} and \textbf{Supplementary information Fig. 8}). The large difference (dissimilarity) between DGN-based CBT and other integration methods-based CBTs can be explained by the fact that DGN, which is based on is a geometric deep learning-based (GDL) architecture, normalizes a population of multigraph networks into a single generic representation in an end-to-end manner unlike other methods except cMGI-Net. Furthermore, the SNL objective introduced in DGN acts as a regularizer to the overfitting and the overlooking of the model while assigning weights to the views, thus helping to avoid view-biased CBT estimation to minimize the distance between the population and its estimated CBT, unlike cMGI-Net. All those different components of DGN architecture promote the generation of distinct CBT, unlike other connectional brain templates. For single-view fusion methods comparison, SM-netFusion outperforms other single-view fusion methods for AD, LMCI, M, and F datasets in both hemispheres (\textbf{Fig~\ref{fig:Hamming-Jaccard}} and \textbf{Supplementary information Fig. 8}).

In summary, our comparative extensive experiments evaluating the performance of the learned brain connectional templates (CBTs) generated by single-view fusion methods and by multigraph integration methods, separately, demonstrate that DGN \citep{gurbuz2020deep} outperforms other multigraph fusion methods in terms of producing the most centered templates, preserving the complex topology of biological networks, and encapsulating the most unique traits of a population of multi-view networks, which makes it easily distinguishable from other reviewed methods. Additionally, we showed that the population-representative connectional template generated by DGN achieves the most similar graph structure with a population of multigraph networks at the local, global, and distance-based scales. DGN proved its efficiency by estimating connectional brain templates that fingerprint the population of multi-view brain networks. For example, DGN-connectional brain templates that fingerprint the population of multi-view brain networks derived from T1-weighted MRI scans have revealed a set of biomarkers for both Alzheimer's diseases and Late Mild Cognitive Impairment.


For single-view integration methods comparison, our experiment results assessing the performance of the generated CBTs demonstrate that SM-NetFusion \citep{mhiri2020supervised} outperforms other single-view fusion methods by producing the most reliable connectional brain templates (CBTs) in terms of centeredness, biomarker discovery of most discriminative brain connections between two populations, and preserving the topological similarity of biological networks on the node-level, global-level, and distance-based level. The CBTs generated by SM-NetFusion capture both most representative and discriminative traits of the multi-view brain networks population while preserving its topological patterns.

\section{Conclusions and future directions}

In this paper, we conducted a primer comparative study of single-view and multi-view graph integration methods for brain connectional template estimation from a population of brain connectomes. We run extensive experiments to evaluate the performance of the reviewed state-of-the-art methods in terms of CBT centeredness, biomarker reproducibility, node-wise similarity, global-level similarity, and distance-based similarity to a ground truth testing dataset. First, we estimated single population-based CBTs and multigraph population-based CBTs by integrating a set of single-view biological networks and a set of multigraph biological networks, respectively into a single connectional template. Next, we computed a set of measurements on the generated CBTs to evaluate their topological properties on the two brain datasets: the Alzheimer's Disease Neuroimaging Initiative (ADNI) database GO public dataset and the Brain Genomics Superstruct Project (GSP) dataset. Based on the experimental results, we demonstrated that SM-netFusion and DGN consistently and significantly outperform other single-view fusion methods and other multigraph integration methods, respectively, by generating well-centered, discriminative, and topologically sound connectional brain templates. Together, these criteria allow SM-netFusion and DGN to lead the discriminative power in discovering connectional biomarkers that disentangle the connectivity variability of two different populations (e.g., healthy vs disordered) of single-view and multi-view brain networks, respectively.

Although SM-netFusion \citep{mhiri2020supervised} outperformed SNF \citep{wang2014similarity} and NAG-FS \citep{mhiri2020joint} in estimating the most reliable CBTs that represent populations of single-view brain networks, it has a few limitations that could be addressed in future work. First, we evaluated SM-netFusion on morphological data for CBTs estimation. In the future work, we aim to explore the discriminative power of brain network atlases derived from other brain modalities such as structural \citep{park2013structural} and functional brain networks. SM-netFusion can be also leveraged to design an efficient feature selection method for training predictive learners in network neuroscience. For example, one can integrate SM-netFusion to boost classification tasks performance by extracting the most discriminative features that differentiate between two graph populations. 

Although DGN \citep{gurbuz2020deep} outperformed other fusion models in the target holistic CBT learning task from heterogeneous brain multigraph datasets, it has a few limitations that could be overcome in future work. First, DGN is limited to static brain networks (fixed-time data points) and may not easily be adapted to more sophisticated network structures such as networks with dynamic connectivity \citep{bessadok2021graph}. Alternatively, multimodal fusion models with flexible and powerful generic architecture can be developed to enable the evaluation of data with a time-dependent brain multigraph population. For instance, geometric recurrent neural networks (RNNs) based on graph convolutional operations can be used to fuse dynamic brain networks derived from MRI measurements acquired at different time points to reveal the trajectory of neurological diseases \citep{ezzine2019learning}. 

Second, all comparative graph fusion methods including DGN, assume that all network views contain the same number of nodes. Alternatively, the DGN model can be extended to handle non-isomorphic graphs with varying numbers of nodes. For instance, \citep{morris2019weisfeiler} propose a generalization of GNNs, so-called k-dimensional GNNs (k-GNNs), which can take higher-order graph structures at multiple scales into account and map any different graphs to different embeddings. Also inspired by GNNs and the Weisfeiler-Lehman (WL) graph isomorphism test, \citep{xu2018powerful} proposed a powerful GNN-based model that distinguishes isomorphic and non-isomorphic graph structures by mapping them to different representations in the embedding space. Recently, \citep{mhiri2021non} proposed the first non-isomorphic inter-modality brain connectome generation GNN architecture. Inspired by such works, DGN can be extended into a non-isomorphic integration GNN network.

Furthermore, DGN is limited by generating single population-based CBT which may not be discriminative enough to disentangle two specific groups. One solution is to train an auxiliary GNN classifier which forces the fusion model to learn how to capture the most discriminative connectional traits differentiating between several brain network populations. Specifically, a classification task can be added to the fusion process to boost the differences between group connectional templates, and thus the learned templates can be useful to study specific population pairs (i.e., brain disorder, gender differences). 

Finally, we evaluated DGN on unimodal data for CBT generation. In future work, we can generalize our comparison study to integrate multi-modal brain networks such as functional and structural brain networks at the same time while capitalizing on geometric deep learning for estimating holistic CBTs and investigating populations differences at functional and structural levels.

The importance of analyzing brain connectivity patterns in biological datasets which proliferate with unprecedented complexity and heterogeneity opens new frontiers to upgrade the capacity of multigraph fusion methods to work on multimodal connectomic datasets to learn integral and holistic connectional templates of populations of multi-view networks. Geometric recurrent neural networks based on CNNs can be used to fuse dynamic graphs acquired at different time points to reveal the trajectory of neurological diseases. Also, multigraph fusion methods can be agnostic to the number of nodes using k-dimensional GNNs-based learning models which map non-isomorphic brain networks to different representations in the embedding space.

\section{Acknowledgements}

This work was funded by generous grants from the European H2020 Marie Sklodowska-Curie action (grant no. 101003403, \url{http://basira-lab.com/normnets/}) to I.R. and the Scientific and Technological Research Council of Turkey to I.R. under the TUBITAK 2232 Fellowship for Outstanding Researchers (no. 118C288, \url{http://basira-lab.com/reprime/}). N.C is also supported by the TUBITAK 2232 Fellowship as a Ph.D student. However, all scientific contributions made in this project are owned and approved solely by the authors. 

\newpage
\bibliography{references}
\bibliographystyle{model2-names}
\end{document}